%% file: main.tex
\documentclass[12pt, reqno]{amsart}

\usepackage{graphics, stackrel}
\usepackage{amsmath, amssymb, amsthm, bm}
\usepackage{graphicx}
\usepackage{verbatim}
\usepackage{amsfonts}
\usepackage{mathtools}

\usepackage{dirtytalk}

\usepackage{caption}
\usepackage{subcaption}
\graphicspath{ {./images/} }

\usepackage[toc,page]{appendix}

\usepackage{tikz}
\usetikzlibrary{arrows}
\usepackage{xcolor}

\usepackage{apacite}
\usepackage{natbib}

\usepackage[ruled, linesnumbered]{algorithm2e}
\SetAlFnt{\small}
\SetAlCapNameFnt{\small}
\SetAlCapFnt{\small}

\usepackage{enumitem}
\setlist[enumerate]{itemsep=2pt,topsep=3pt}
\setlist[itemize]{itemsep=2pt,topsep=3pt}
\setlist[enumerate,1]{label=(\alph*)}

\usepackage{booktabs}

\usepackage[T1]{fontenc}
\usepackage{dsfont}

\usepackage{mathrsfs}
\usepackage{bbm}  

\renewcommand{\phi}{\varphi}
\renewcommand{\epsilon}{\varepsilon}

\usepackage[left=1.25in, right=1.25in, top=1.0in, bottom=1.15in, includehead, includefoot]{geometry}
\usepackage[citecolor=blue, colorlinks=true, urlcolor=blue, linkcolor=blue]{hyperref}

\definecolor{darkpowderblue}{rgb}{0.0, 0.2, 0.6}
\definecolor{indigo}{rgb}{0.0, 0.25, 0.42}
\definecolor{charcoal}{rgb}{0.21, 0.27, 0.31}
\definecolor{royalblue}{rgb}{0.0, 0.14, 0.4}

\usepackage{hyperref}  
\hypersetup{
    colorlinks=true,
    linkcolor=indigo,
    citecolor=royalblue,
    filecolor=magenta,      
    urlcolor=black,
}

\renewcommand{\leq}{\leqslant}
\renewcommand{\geq}{\geqslant}

\DeclareMathOperator{\diag}{diag}  

\DeclareMathOperator{\Prob}{Prob}

\newcommand{\rom}[1]{\lowercase\expandafter{\romannumeral #1\relax}}

\newcommand{\de}{\,{\rm d}} 

\theoremstyle{plain}
\newtheorem{theorem}{Theorem}[section]
\newtheorem{corollary}[theorem]{Corollary}
\newtheorem{lemma}[theorem]{Lemma}
\newtheorem{proposition}[theorem]{Proposition}

\theoremstyle{definition}
\newtheorem{definition}{Definition}[section]

\newtheorem{example}{Example}[section]
\newtheorem{remark}{Remark}[section]

\newtheorem{assumption}{Assumption}[section]

\def\R{{\mathbbm R}}
\def\Q{{\mathbbm Q}}

\def\N{{\mathbbm N}}

\def\E{{\mathbbm E}}
\def\p{\partial}

\def\a{\alpha}
\def\b{\beta}
\def\g{\gamma}

\def\d{\delta}

\def\e{\epsilon}

\def\l{\lambda}

\def\k{\kappa}
\def\s{\sigma}
\def\th{\theta}

\def\lp{\left(}
\def\rp{\right)}
\def\lb{\left[}
\def\rb{\right]}

\def\diag{\mathrm{diag}}

\def\graph{\mathrm{graph\,}}

\def\diag{\mathrm{diag\,}}

\def\ra{\rightarrow}

\makeatletter
\g@addto@macro{\endabstract}{\@setabstract}
\newcommand{\authorfootnotes}{\renewcommand\thefootnote{\@fnsymbol\c@footnote}}%
\makeatother

\begin{document}
\begin{center}
  \Large{Uniqueness of Equilibria in Interactive Networks}

    \vspace{1em}

  \normalsize
  \authorfootnotes
  Chien-Hsiang Yeh\footnote{Email: \href{mailto:chien.yeh@anu.edu.au}{chien.yeh@anu.edu.au}. The
  author thanks John Stachurski for his instruction and suggestion.
  The author also thanks Timothy Kam for his invaluable comment.
  }

  \par \bigskip

  \small
  Research School of Economics, Australian National University \bigskip

  \normalsize
  \today
\end{center}

\begin{abstract}
\vspace{1em}
This paper extends the unified network model, proposed by \cite{acemoglu2016network}, such that interaction functions can be heterogeneous, and the sensitivity matrix has less than or equal to one spectral radius.
We show the existence and (almost surely) uniqueness of equilibrium under both eventually contracting and non-contracting assumptions.
Applying the equilibrium in the study of systemic risk, we provide a measure to determine the key player who causes the most significant impact if removed from the network. 

\vspace{1em}
    \noindent
    \textit{Keywords:} Network, Equilibrium, Uniqueness,  Key Player
\end{abstract}

\setcounter{footnote}{0}

\input{introduction}

\input{model}

\input{uniqueness}

\vspace{20mm}
\input{appendix}

\bibliographystyle{ecta}
\bibliography{main}


\end{document}

%% file: introduction.tex
\section{Introduction}

In the last three decades, since network models capture the interaction among agents, the study of networks helps us understand macroeconomic volatility and systemic risk.
For instance, it is well-known that the idiosyncratic shocks can propagate through the interconnections of a production network and then leads to aggregate fluctuation.
The structure of network graphs, such as hubs, sparsity, and asymmetry structure, influences the shock propagation and the magnitude of aggregate fluctuation \citep{production:carvalho2008, Production:acemoglu2012network, production:carvalho2014micro, production:carvalho2019production}.


In the study of networks, it is significant to determine whether the equilibrium is unique, since it may have different economic properties and interpretations, and the comparative statics may also fail if multiplicity exists.
For example, \cite{multiplicity:roukny2018interconnectedness} address that the multiple equilibria of a financial credit network make the probability of default indeterminate and then make it challenging to evaluate systemic risk.
\cite{multiplicity:jackson2020credit} indicate that the multiple equilibria of a financial network lead to a self-fulfilling cascade of defaults.\footnote{When there are multiple equilibria and the market has pessimistic beliefs, a bank may tend to hold cash and stop payments to others if it believes that other banks experience deterioration in credit conditions.
In this case, the ex-ante fear causes the cascade of defaults, even there is another better equilibrium that banks are solvent.}
Therefore, it is necessary to check the uniqueness of equilibrium in the network models.
To this end, this paper attempts to find the conditions for the uniqueness of equilibrium.


We study the question: given a list of different network models, is there a unified framework to check the uniqueness easily and quickly? 
To answer the question, this paper generalizes the unified framework proposed by \cite{acemoglu2016network} to embody more network models, so that it is more likely to be used in the future research.
The main contribution of this paper is that we show the existence and (almost surely) uniqueness of equilibrium under general and commonly used assumptions, so that it is easier to apply our results to check (almost surely) uniqueness.

In detail, this paper considers the heterogeneous interaction functions and arbitrary sensitivity matrix in the network model of \cite{acemoglu2016network}.
As a result, the unified network model embodies the financial networks such as \cite{Financial:liu2020interbank}, who simulate the U.S. interbank lending system to study the contagion effect of bank failures and confirm that the extent of network contagion effect has been reduced after the 2007-09 financial crisis.\footnote{They find that banks have fewer counterparty exposures after the financial crisis.} 
They only show the existence of equilibrium.
Our result shows that the clearing payment for their generalized Eisenberg-Noe model is uniqueness almost surely.\footnote{We also show the uniqueness of clearing payments for Eisenberg-Noe model without assuming regularity.}
Hence, when the shocks are absolutely continuous, the probability of multiple clearing payments is zero, so that their simulation is well-behaved.

We show that the equilibrium exists and is unique when either the interaction function or the sensitivity matrix is convergent such that there exists Banach contraction. 
That is, if the spectral radius of the sensitivity matrix weighted by the Lipschitz constants of interaction functions is less than one, then the equilibrium is unique and globally stable.  
Hence, this result is different from \cite{acemoglu2016network} that the uniqueness of equilibrium does not merely depend on the Lipschitz contraction of the interaction functions.
We also check the tightness of assumptions and find that the assumptions are necessary if we require the economic states to be positive.

On the other hand, we show that equilibrium exists and is unique almost surely, if the interaction functions are non-expansive but bounded, the sensitivity matrix is non-convergent (i.e., spectral radius one.), and the shocks are absolutely continuous.
We notice that the absolute continuity is essential to prevent the confusion of non-zero probability of multiplicity.
For instance, multiple equilibria may occur with strictly positive probability in some financial networks, such as \cite{acemoglu2015systemic} and \cite{acemoglu2015Endogenous}.\footnote{The generic uniqueness is defined in the sense that the set of shocks admitting multiplicity is measure zero.}

Furthermore, study the non-bounded linear system and argue that the boundedness condition is essential to both the existence and uniqueness of equilibrium, when the network is non-contracting.
We use this concept to design an algorithm to compute the generalized Eisenberg-Noe interbank lending model in \cite{acemoglu2015systemic} and \cite{acemoglu2015Endogenous}, where the interaction functions are bounded identity maps.
We show that the algorithm converges in at most $n 2^{n-1}$ iterations.
Moreover, as a special case of the methodology, we also show that the equilibrium payment in \cite{eisenberg2001systemic} of credit network is always unique without regularity condition. 
This result is as the same as \cite{stachurski2022systemic}.

For the application of the generalized network and the unique equilibrium, we further provide the measure for identifying the key players.
We follow the methodology in \cite{sharkey2017control} and generalize the measure to our model. 
The \emph{key players} are the agents who create the highest reduction of aggregate economic states if they are removed from the network \citep{networkgame:ballester2006s, NetworkGame:zenou2016key}.
Using the methodology of \cite{sharkey2017control}, we calculate the measure by interpreting the equilibrium as the steady state of a continuous-time dynamic system.
The benefit of the measure is that it captures the impact of both the received shocks from others and the shocks that agents pass on.
About systemic risk, the measure also has the feature that the identified key players are either too-big-to-fail or too-interconnected-to-fail agents.

\subsubsection*{Related Literature}

The unified network model in this paper can be applied to determine the Nash equilibrium in network games, the equilibrium output in input-output analysis, and the clearing payments in generalized Eisenberg-Noe financial networks \citep{eisenberg2001systemic}.
For example, the model can be used to describe the best response and solve Nash equilibrium in network games  \citep{networkgame:calvo2009peer, networkgame:cohen2015static, networkgame:blume2015linear, NetworkGame:zenou2016key, networkgame:galeotti2020targeting}.
For production networks, the model could represent the input-output relationship and determine the output equilibrium  \citep{Production:acemoglu2012network, production:Bartelme_2015, Production:acemoglu_Akcigit_Kerr_2016, production:acemoglu2017microeconomic, production:herskovic2018networks, production:carvalho2019production, production:acemoglu2020endogenous, production:herskovic2020firm, production:pesaran2020econometric}.
For financial networks, it calculates the clearing loan repayments, which studies the systemic risk of default cascade \citep{eisenberg2001systemic, Financial:cifuentes2005liquidity, financial:elsinger2006risk, financial:rogers2013failure, Financial:glasserman2015likely, financial:glasserman2016contagion, acemoglu2015systemic, Financial:gai2019networks, Financial:veraart2020distress}.



\subsubsection*{Outline} The paper is organized as follows.
Section~\ref{sec:model and example} presents the unified model that builds in all networks in Section~\ref{sec:NetworkModels}, which lists out the network models with identical mathematical patterns.
Section~\ref{sec:generic uniqueness} shows the existence and (almost surely) uniqueness of equilibrium.
Section~\ref{sec:tightness and boundedness} presents the comparative statics, and discusses the tightness of assumptions and the importance of boundedness condition.
Section~\ref{sec:algorithm} provides an algorithm to compute the equilibrium when the interaction functions are bounded identity maps.
Section~\ref{sec:key player} presents a measure for identifying key players, which utilizes the property of the unique equilibrium.

%% file: model.tex
\section{General Model} 
\label{sec:model and example}

\subsection*{Notation and Preliminary} \label{section:preliminaries}
Let $x, y \in \R^n$ be vectors and $f_i:\R \ra \R$ be functions for all $i\in V\coloneqq \{ 1,\dots, n\}$. In expressions involving matrix algebra, we take the
convention that all vectors are row vectors, unless otherwise stated.
Denote $|x|$ as $|x|\coloneqq (|x_1|, \dots, |x_n|)$ and $f(x)$ as $f(x)\coloneqq (f_1(x_1),  \dots, f_n(x_n))$. 
If $f_i \equiv f$ for all $i \in V$, we write $f(x)\coloneqq (f(x_1),  \dots, f(x_n))$.
We say that $f_i$ is \emph{non-expansive} if $|f_i(x) - f_i(y)| \leq | x - y|$ for all $x, y \in \R$.
Also, $f=(f_i)_{i\in V}$ is \emph{non-expansive} if $f_i$ is non-expansive for all $i$.
Denote $x \geq y$ if $x_i \geq y_i$ for all $i \in V$, $x > y$ if $x_i \geq y_i$ for all $i$ and $x_i > y_i$ for some $i\in V$, and $x \gg y$ if $x_i > y_i$ for all $i\in V$.

Let $A=(A_{ij}), B=(B_{ij}) \in \R^{n\times n}$ be square matrices.
Similarly, denote $A \geq B$ if $A_{ij} \geq  B_{ij}$ for all $i,j \in V$ and $A \gg B$ if $A_{ij} > B_{ij}$ for all $i,j \in V$.
A matrix $A$ is \emph{column (row) stochastic} if $A \geq 0$ and $\sum_{i\in V} A_{ij} = 1$ for all $j \in V$ ($\sum_{j \in V} A_{ij} = 1$ for all $i \in V$).
A matrix $A$ is \emph{column (row) substochastic} if $A \geq 0$ and $\sum_{i\in V} A_{ij} \leq 1$  for all $j \in V$ ($\sum_{j \in V} A_{ij} \leq 1$ for all $i \in V$).
A matrix $A$ is \emph{irreducible} if $\sum_k A^k \gg 0$.
The norm $\| \cdot \|$ refers to $p$-norm for vectors or matrix norm induced by $p$-norm.
Denote $r(A)\coloneqq \{\max |\mu|:\text{$\mu$ is an eigenvalue of $A$} \}$ as the \emph{spectral radius} of $A$.
A matrix $A$ is \emph{convergent} if $\lim_{k \ra \infty } (A^k)_{ij} = 0$ for all $i, j\in V$, where $(A^k)_{ij}$ is the $(i,j)$-th entry of $A^k$.
Let $x\in \R^n$ be a vector. We write $\diag(x)$ as the diagonal matrix with main diagonal $x$.
We know that matrix $A$ is convergent if and only if $\lim_{k \ra \infty} \| A^k\| = 0$ or $r(A) < 1$. We say that $W$ is \emph{weakly chained substochastic} if $W$ is row substochastic and for each $i \in N$, either $\sum_j w_{ij} < 1$, or there exists $t \in N$ such that there is a path from $i$ to $t$ ($i \ra \cdots \ra t$) and $\sum_j w_{tj}<1$.
We can check the convergence by weakly chained substochasticity (see \cite{azimzadeh2019contraction}.)
\begin{lemma}\label{condition:convergence}
If $W \geq 0$ or $W^\top \geq0$ is weakly chained substochastic, then $W$ is convergent.\footnote{$W^\top$ is weakly chained substochastic if and only if $W$ is column substochastic, and for each $j \in N$, either $\sum_i w_{ij} < 1$, or there exists $t \in N$ such that there is a path from $t$ to $j$ ($t \ra \cdots \ra j$) and $\sum_i w_{it} < 1$.}
\end{lemma}

Given a square matrix $A = (a_{ij}) \in \R^{n\times n}$, \emph{graph $A$} is a tuple $(V, E)$ consisting of the vertex set $V\coloneqq \{1, \dots, n\}$  and the edge set $E \subset V\times V$ set such that $(i, j) \in E$ if and only if $a_{ij} \neq 0$.
The $\graph A$ has a \emph{directed path} $i \ra j$ for $i, j \in V$ if $a^k_{ij} > 0$ for some $k \in \N$, where $a^k_{ij}$ is the $(i,j)$th entry of $A^k$.
A $\graph S = (V_S, E_S)$ is a \emph{subgraph} of $\graph A$ if $V_S \subset V$ and $E_S \subset E$.
In the following model, $a_{ij} \neq 0$ means agent $i$'s state affects agent $j$'s state or $j$'s state is sensitive to $i$'s state.
A vertex $j$ is \emph{accessible} from a vertex $i$ if either $i=j$ or there is a directed path $i \ra j$.
Graph $A$ is \emph{strongly connected} if vertex $j$ is accessible from vertex $i$ for any $i, j \in V$.
We know that graph $A$ is strongly connected if and only if $A$ is irreducible.
We say a $\graph A$ is \emph{acyclic} if $i$ is not accessible from $j$ whenever $j$ is accessible from $i$ for all $i\neq j \in V$.
Write $\bm 1 \in \R^n$ as the vector of ones and $\mathbbm{1}_A$ as the indicator function given a set $A$ such that $\mathbbm{1}_A(z) = 1$ if $z \in A$ and $\mathbbm{1}_A(z) = 0$ otherwise.
Denote $\lambda$ as the Lebesgue measure.


\subsection{Model}
Consider an economy with $n \geq 2$ agents, indexed by $N = \{1, \dots, n\}$.
Each agent's economic state is $x_i \in \R$.
Agent $j$'s state depends on the other agents' states:
\begin{equation} \label{eq:interation function}
x_j = f_j \left(\sum_{i=1}^n x_i w_{ij}  + \epsilon_j \right)
\end{equation}
where $f_j\colon \R \ra \R$ is called the \emph{interaction function} which describes how shocks and other agents' states affect agent $j$, $w_{ij}$ is the sensitivity extent of interaction between $i$ and $j$, and $\e_j$ is a shock.
We call $W\coloneqq (w_{ij}) \in \R^{n \times n}$ as the \emph{sensitivity matrix},  whose entries are $w_{ij}$.
Let $x \coloneqq ( x_1, \dots, x_n)$, $\e\coloneqq(\e_1, \dots, \e_n )$, and 
$$f(xW+\e)  \coloneqq \lp f_1 \lp\sum_{i} x_i w_{i1} + \e_1 \rp, \dots, f_n \lp\sum_{i} x_i w_{in} + \e_n \rp \rp.$$
Equation (\ref{eq:interation function}) can be rewritten in the vector form:
\begin{equation*}
    x = f(xW + \e).
\end{equation*}

We call $(f, W)$ a \emph{network}. 
Also, we let $\mathbbm{P}_\e$ be the probability distribution of $\e$.
As presented in the subsequent section, the economic states can be the outputs in production networks, the decision choice in network games, and the amount of borrowing in financial networks.
The model (\ref{eq:interation function}) indicates that agent $i$'s state influences agent $j$'s state if and only if $w_{ij} \neq 0$.
The interaction links and network structure are presented by $\graph W$.
The extent of influence from other agents is decided by both the sensitivity $w_{ij}$ and interaction function $f_j$.
Figure \ref{fig:network model} plots an example network, where only the paths with non-zero sensitivity $w_{ij}$ are presented.
In this case, the equilibrium state of the agent $3$ depends on the states of all the other agents.
But, agent $4$'s and $5$'s states also depend on agent $3$'s state, so there exists feedback loop to agent $3$ from agent $4$ and $5$. 
We see that the network model is complex, even there are only $5$ agents in the network.

\begin{figure}
    \centering
    \begin{tikzpicture}[baseline, auto, node distance={20mm}, thick, main/.style = {draw, circle, scale=0.8},] 
        \node[draw, circle, label={$\e_1$}] (1) at (0, 0) {$1$};
        \node[draw, circle, label={$\e_2$}] (2) at (3, 0) {$2$};
        \node[draw, circle, label={$\e_3$}] (3) at (6, 1) {$3$};
        \node[draw, circle, label=below:{$\e_4$}] (4) at (6, -1) {$4$};
        \node[draw, circle, label=right:{$\e_5$}] (5) at (9, 0) {$5$};

        \path [->,bend left=10, above] (1) edge node {$w_{12}$} (2);
        \path [->,bend left=10, below] (2) edge node {$w_{21}$} (1);
        \path [->,bend left=10, above] (2) edge node {$w_{23}$} (3);
        \path [->,bend left=-10, below] (2) edge node {$w_{24}$} (4);
        \path [->,bend left=10, right] (3) edge node {$w_{34}$} (4);
        \path [->,bend left=10, left] (4) edge node {$w_{43}$} (3);
        \path [->,bend left=10, above] (3) edge node {$w_{35}$} (5);
        \path [->,bend left=10, below] (5) edge node {$w_{54}$} (4);
        \end{tikzpicture}
    \caption{}
    \label{fig:network model}
\end{figure}

Note that \cite{acemoglu2016network} assume that all agents have the identical interaction function $f_j \equiv f$ for all $j \in N$, while equation (\ref{eq:interation function}) can have heterogeneous interaction functions so that it includes more network models.
We list out the networks embodied in the interaction model (\ref{eq:interation function}) in the following subsection.

\input{networks}

\subsection{Equilibrium}\label{sec:equilibrium}
Following \cite{acemoglu2016network}, the equilibrium is defined as follows.

\begin{definition}
Given the realization of the shocks $(\e_1, \dots, \e_n)$, an \emph{equilibrium} of the economy is a collection of states $(x_1, \dots, x_n)$ such that equation (\ref{eq:interation function}) holds for all agents simultaneously.
\end{definition}

In other words, the equilibrium is a vector of values $x=(x_i)_i$ that solves $x = f(xW + \e)$ given the $\graph W$ and shock $\e$.
Next, since a network may exist multiple equilibria, we define the "almost sure uniqueness of equilibrium".

\begin{definition} \label{def:almost surely unique}
Let $E \subset \R^n$ denote the set of $\e$. 
Denote the subset $M$ of $E$ as  $M \coloneqq \{\e \in E: \text{Equation (\ref{eq:interation function}) has multiple equilibria} \}$.
A network has \emph{almost surely unique} equilibrium, or \emph{almost sure uniqueness of equilibrium} holds, if the equilibrium exists for $\e \in E\setminus M$ and $\mathbbm{P}_\e(\e \in M ) = 0$, where $\mathbbm{P}_\e$ is the distribution of $\e$.\footnote{Let $(\Omega, \mathcal{F}, \mathbbm{P})$ be the measure space. Distribution $\mathbbm{P}_\e(B) \coloneqq \mathbbm{P}(\e^{-1}(B))$ for all $B \in \mathcal{B}(\R)$.}
\end{definition}

We say that the equilibrium is \emph{unique almost surely} when the network satisfies Definition \ref{def:almost surely unique}.
When almost sure uniqueness of equilibrium holds, the probability that the shock admits multiple equilibria is zero.
Therefore, as we will show, the distribution of shock determines whether there are multiple equilibria or not.
If the probability distribution of shock is discrete, then the multiple equilibria may occur with non-zero probability.
We do not use the terminology of generic uniqueness, defined in \cite{acemoglu2016network} and the following remark, to exclude the problematic cases that admit multiple equilibria with strictly positive probability.

\begin{remark}
In this remark, we first list out the assumptions on $(f, W, \e)$ in \cite{acemoglu2016network} for comparison and then discuss the generic uniqueness and its confusing result. 
\cite{acemoglu2016network} consider the model with homogeneous interaction functions $f_i \equiv g$.
They assume that $g$ is continuous, increasing, and either contracting or non-expansive but bounded, $W$ is column stochastic, and the shocks $\e_i$ are independently and identically distributed with mean zero and constant variance.\footnote{Since $W$ is column stochastic, every agent has constant total dependence on others.}
According to \cite{acemoglu2016network}, we define the generic uniqueness of equilibrium as follows.\footnote{\cite{acemoglu2016network} explain the generic uniqueness as \say{\textit{\dots is generically unique, in the sense that the economy has multiple equilibria only for a measure zero set of realizations of agents-level shocks.}}
}

\begin{definition}\label{def:generic uniqueness:Lebesgue}
Let $D \subset \R^n$ denote the set of $\e$.
Let $M$ denote the set of shocks that admit multiple equilibria (i.e., $M \coloneqq \{\e \in D: \text{Equation (\ref{eq:interation function}) has multiple equilibria} \}$.)
\emph{Generic uniqueness} holds if the equilibrium exists for all $\e \in D\setminus M$, and the Lebesgue measure of $M$ is zero, $\lambda(M) = 0$.
\end{definition}

Under the above assumptions, \cite{acemoglu2016network} show the existence and generic uniqueness of equilibrium for any networks.\footnote{
\cite{acemoglu2016network} show the generic uniqueness with the assumption of strong connectedness. 
They do not provide proof for extending the strongly connected graph to a general graph.
We extend their proof to any network in Section \ref{sec:generic uniqueness}.}
While their result is true, it may cause some confusion, since measure-zero events can also be probability-one events.
We show in Example \ref{example:counter} that if the shock variable is discrete, there may be high probability of multiplicity of equilibria.
Therefore, if we follow Definition \ref{def:generic uniqueness:Lebesgue}, then depending on the shock specification, it might be the case that generic uniqueness holds, but the multiplicity of equilibria exists with arbitrarily high probability. 
Since this causes confusion, we adopt Definition \ref{def:almost surely unique}.

Furthermore, the above assumptions of \cite{acemoglu2016network} do not nest the networks (\ref{eq:input-output}), (\ref{eq:networkgame:best-reply}) - (\ref{eq:EN model:bankrupty cost}), since they require that the sensitivity matrix is column stochastic, and
the interaction functions are identical $f_i \equiv f$.
For instance, network (\ref{eq:generalised EN model}) has row stochastic sensitivity matrix and heterogeneous interaction functions, given that $\bar p_i$ is not identical for all $i$. 
To this end, we show the existence and almost sure uniqueness under more general assumptions embodying more network models.
\qed
\end{remark}

%% file: networks.tex
\subsection{Network Examples} \label{sec:NetworkModels}
This section quotes some network models about production networks (Section \ref{section:InputOurput} and \ref{section:production}), network games (Section \ref{section:SimpleNetworkGame} and \ref{section:NetworkGame})  and financial networks (Section \ref{section:CrossHoldings}, \ref{section:Financial}, \ref{section:Financial:BankruptcyCost} and \ref{section:financial:Liu et al}). 
It aims to show that these network models have the same mathematical pattern as equation (\ref{eq:interation function}).

Moreover, these examples illustrate that the interaction functions and sensitivity matrix follow similar mathematical properties. 
In particular, most of the interaction functions are monotone and Lipschitz continuous.
We also list out the conditions for the existence and uniqueness of equilibrium for each model.\footnote{The equilibrium is explicitly defined in Section \ref{sec:equilibrium}.}

 
\subsubsection{Input-Output Analysis}\label{section:InputOurput}
The input-output analysis describes the inter-industry relationship by a matrix that tracking the flow the money.\footnote{See \cite{miller2009input} and \cite{miller2017Downstreamness_Upstreamness} for example.}
It studies how the shock in one sector affects the other sectors' output.\footnote{\cite{InputOutput:fletcher1989tourism} uses input-output analysis to study the impact of tourism.
}
The analysis also helps to identify which industry or region is the most significant to optimize aggregate economy,
The model is briefly introduced as below.

There are $n$ industries in a closed economy with no inventories.
Each industry $i \in \{1, \dots, n\}$ requires $w_{ij} \in [0, 1]$ dollar amount of intermediate input from industry $j \in \{1, \dots, n\}$ to produce one dollar of $i's$ output.
Input-output tables determine the linkage weights between sectors $w_{ij}$ empirically.\footnote{See \cite{InputOutputTable:timmer2015}.} 
For every industry $j \in \{1, \dots, n\}$, the gross output $x_j$ equals the total value of its use as a final good $\e_j$ and its use as an intermediate input to other industries:
\begin{equation} \label{eq:input-output}
    x_j = \e_j + \sum_i x_i w_{ij}.
\end{equation}
That is, the sale of industry $j$ to other sectors is $\sum_i x_i w_{ij}$.
To compute the equilibrium, it is conventional to assume that every sector has some inputs from labor or other value-added, so that $\sum_{j} w_{ij} < 1$ for all $i$.
Let $x = (x_i),\, \e = (\e_i) \in \R^n$ and $W = (w_{ij}) \in \R^{n \times n}$.
The vector form of equation (\ref{eq:input-output}) is $x = \e + x W$.
Since the row sum of $W$ is less than one, the matrix $(I-W)$ is non-singular.
The unique equilibrium of output is $x = \e(I-W)^{-1} $, where $(I-W)^{-1}$ is known as Leontief inverse.

\subsubsection{Production Networks}\label{section:production}
This subsection presents the production network  of \cite{production:long1983real}, \cite{production:carvalho2008}, \cite{Production:acemoglu2012network},
\cite{Production:acemoglu_Akcigit_Kerr_2016},
\cite{production:acemoglu2017microeconomic}, \cite{production:carvalho2019production} and \cite{production:acemoglu2020endogenous}.
The production network investigates how the heterogeneous shock to an individual sector can generate aggregate fluctuations, given the supplier-customer interconnections in a production network.
Moreover, the network model illustrates that such aggregate fluctuation and the cascade effect of shocks are correlated with the structure of the networks (\cite{production:carvalho2008} and \cite{Production:acemoglu2012network}). 

The economy has $n$ competitive sectors. 
Each sector's output $x_j$ follows the production function:
\begin{equation*}
    y_j = z_j^\alpha \ell_j^\alpha \prod_{i=1}^n y_{ij}^{(1-\a) w_{ij}}
\end{equation*}
where $z_j$ denotes the productivity shock, $\ell_j$ denotes the labor input, $\a \in (0, 1)$ is the share of labor, $y_{ij}$ is the intermediate input from sector $i$ used in the production of good $j$, and $w_{ij} \geq 0$ is the share of intermediate input $i$ in the total intermediate input.
It is supposed that $\sum_i w_{ij} = 1$ for all $j=1, \dots, n$.
Let $p_j$ be the price of good $j$ and $h$ be the labor wage. 
Producers maximize their profits:
\begin{align*}
    \max_{\ell_j, y_{1j}, \dots, y_{nj}} p_j y_j - h \ell_j - \sum_{i=1}^n p_i y_{ij} 
\end{align*}
The optimal labor input is $\ell_j = \a p_j y_j / h$ and the intermediate input is $y_{ij} = (1-\a)  w_{ij} p_j y_j / p_i$.
The representative household with Cobb-Douglas preferences solves the optimal problem:
\begin{equation*}
    \begin{split}
        \max_{c_1, \dots, c_n} u(c_1, \dots, c_n) = A \prod_{j=1}^n c_j^{1/n} \qquad \text{s.t. } \sum_j p_j c_j = h \sum_j \ell_j 
    \end{split}
\end{equation*}
where $c_j$ is the consumption of good $h$, and $A$ is a normalization constant.
Normalizing the total labor supply $\sum_i \ell_i=1$, the first order condition of the optimal consumption gives $c_j = h / (n p_j)$.
The clearing condition of commodity market, from equation (\ref{eq:input-output}), is $y_j = c_j + \sum_k y_{jk}$.
Then, we have 
\begin{equation} \label{eq:production_clearing}
   p_j y_j = \frac{h}{n} + (1-\a) \sum_{k=1}^n w_{jk} p_k y_k  
\end{equation}
Let $\hat y_j = p_j y_j$, $\hat y = (\hat y_j)\in \R^n$ and $W = (w_{ij})\in \R^{n\times n}$.
The vector form of clearing condition is $\hat y = (h/n) \bm 1 + (1-\a) \hat y W^\top$.
 Since the row sum of $(1 - \a) W^{\top}$ is strictly less than one, $[I - (1 - \a) W^{\top}]$ is non-singular.
Thus, we have is $\hat y = (h/n) \bm 1 [1-(1-\a) W^\top]^{-1} $.
Defining $b \coloneqq  \bm 1 [I - (1 - \a) W^{\top}]^{-1}$, we write $p_j y_j = b_j h / n$.
Hence, it yields $\ell_j = \a b_j/n$ and $y_{ij} = (1-\a) w_{ij} y_i  b_j / b_i $, so the production function gives 
\begin{equation*}
\log y_j = \mu_j + \a \log z_j + (1-\a) \sum_i (\log y_i) w_{ij}     
\end{equation*}
where $\mu_j$ is some constant.\footnote{$\mu_j = \log(b_j (\a/n)^\a (1-\a)^{1-\a}) + (1-\a) \sum_i w_{ij} \log( w_{ij} / b_i)$.}
Denote $x_j = \log y_j$ and $\e_j = (\mu_j + \a \log z_j)/(1-\a)$, it delivers
\begin{equation} \label{eq:production:log output}
    x_j = (1-\a) \lp \sum_{i} x_i w_{ij} +  \e_j \rp
\end{equation}
The equilibrium output is $x = \e[I - (1-\a)W]^{-1} $.
Overall, there are two simple network equations (\ref{eq:production_clearing}) and (\ref{eq:production:log output}).

\subsubsection{Simple Network Games}\label{section:SimpleNetworkGame}
In a social network game, an agent's payoff or well-being not only depends on her action, but also depends on her neighbors' actions.
Social network influences decision behavior, such as committing a crime and lending decision (\cite{NetworkGame:ballester2004s:crime} and \cite{networkgame:cohen2015static}).
Consider a simple network game as \cite{networkgame:zenou2012networks}, \cite{NetworkGame:zenou2016key} and \cite{networkgame:galeotti2020targeting}.
There are $n$ players in a social network, and the social connection is represented by $\graph W$.
If agent $i$ is connected with agent $j$, then $w_{ij}=1$; otherwise, $w_{ij}=0$.
Moreover, assume that $w_{ii}=0$ for all $i=1, \dots, n$ by convention.
Thus, $W=(w_{ij})$ is an adjacency matrix with entry $w_{ij}$.
Assume that it is a game of strategic complement with perfect information such that players know everything about the network.
Agents choose the actions $x_j \in \R_+$ to maximize their payoffs:
$$u_j(x_1, \dots, x_n) = \a_j x_j - \frac{1}{2}x_j^2 + \phi \sum_{i=1}^n w_{ij} x_i x_j $$
where $\a_j > 0 $ is the exogenous heterogeneity capturing individual characteristics, $\a_j x_j - (1/2)x_j^2$ is the individual benefits, and $\phi \sum_{i=1}^n w_{ij} x_i x_j$ is the peer influence depends on the location of agents.
Hence, every agent's payoff depends on her own action and the other agents' actions.
The best-reply function in equilibrium is 
\begin{equation}\label{eq:networkgame:best-reply}
    x_j = \a_j + \phi \sum_{i=1}^n x_i w_{ij} = \phi \lp  \sum_{i=1}^n x_i w_{ij} + \e_j \rp
\end{equation}
where we let $\e_j = \a_j/\phi$.
For the equilibrium, suppose that $\phi \, r(W) < 1$ so that $\phi W$ is non-singular.
The unique Nash equilibrium is $x^* = \e (I - \phi W)^{-1} $.
Note that we call $\e_j$ as the shock in our model, but in network game it captures the observable characteristics of individual $j$ such that it is exogenous.

\subsubsection{Network Games with Global and Local Interaction}\label{section:NetworkGame}
\cite{networkgame:ballester2006s} consider both the global substitutability and local influence complementarity in network games. 
They investigate how to identify the "key player" that, once removed, causes the maximal decrease in aggregate activity.
The model is similar to the simple network game in the previous section.
Let $G = (g_{ij})$ be the adjacency matrix such that $g_{ii}=0$ for all $i=1, \dots, n$, and $g_{ij} = 1$ if $i$ and $j$ are connected and $g_{ij}=0$ otherwise.
Following the setup in the previous section, given the action profile $(x_i)$, each agent $j$ have the alternative payoff:
$$ u_j(x_1, \dots, x_n) = \a_j x_j - \frac{1}{2} (\eta - \g) x_j^2 - \g \sum_{i=1}^n x_{i} x_j + \phi \sum_{i=1}^n g_{ij} x_i x_j $$
where $\a_j > 0$ for all $j$, $\eta, \phi >0$, $\g \geq 0$, and $\g \sum_{i=1}^n x_{i} x_j$ denotes the global interaction of substitute effect across all agents, and the last term represents the local interaction of strategic complement as the before.
The best-reply function is:
\begin{equation} \label{eq:networkgame:best-reply:general}
    x_j = \frac{\a_j}{\eta } - \frac{\g}{\eta} \sum_{i=1}^n x_i + \frac{\phi}{\eta}\sum_{i=1}^n x_i g_{ij} = \frac{\phi}{\eta} \lp  \sum_{i=1}^n x_i w_{ij} + \e_j \rp
\end{equation}
where $w_{ij} = g_{ij} - \g/\phi$ and $\e_j = \a_j / \phi $.
We see that the best-reply function (\ref{eq:networkgame:best-reply:general}) has the same form as (\ref{eq:interation function}).

Denote $J \in \R^{n \times n}$ as the square matrix of ones.
The Nash equilibrium solves $\eta x^* = \a  - \g x^* J + \phi x^* G$, where $\a = (\a_i)$.
If $(\eta I + \g J - \phi G)$ or $(\phi / \eta) W$ is non-singular,
then the Nash equilibrium is $\a (\eta I + \g J - \phi G)^{-1} $.\footnote{
We can further solve the Nash equilibrium. Suppose that $W$ is non-singular. Then $x^*$ is unique and $\overline{x}^* \coloneqq \sum_j x^*_j$ is constant, 
We further assume that $\phi r(G) < \eta$.
Since $x^* J = \overline{x}^* \bm 1$, we also have $x^* = (\a - \g \, \overline x^* \bm 1) (\eta I - \phi G) ^{-1}$.
Then, we can find that $\overline{x}^* = \overline{b}_\a / (1 + \g \overline b)$, where $b_\a = \a (\eta I - \phi G)^{-1} $, $b = \bm 1(\eta I - \phi G)^{-1}$, $\overline b_\a = b_\a \bm 1^{\top}$, and $\overline b = b \bm 1^{\top}$.
The Nash equilibrium is $x^* = b_\a - \g [\overline b_\a / (1 + \g \overline b)] b $.
}

For example, \cite{networkgame:cohen2015static} present an interbank lending network with the same features.
They consider a network with $n$ banks in a lending market, and the adjacency matrix is that $g_{ij} = 1$ if bank $j$ makes a loan to bank $i$ and $g_{ij}=0$ otherwise.
Each bank $j$ has the profit function given its volume of loans $x_j$ to other banks:
\begin{equation*} 
\pi_j = p x_j - c_j x_j = \lp \th - \sum_{i=1}^n x_j \rp x_j - \lp c_{0,j} - \phi_j \sum_{i=1}^n x_i g_{ij} \rp x_j  
\end{equation*}
where the $p = \th - \sum_{i=1}^n x_i $ is price of loans determining the interest rate, and $c_j =  c_{0,j} - \phi_j \sum_{i=1}^n x_i g_{ij}$ is the marginal cost, with $\th > 0$ and $ c_{0,j}, \phi_j > 0$ for all $j$.
The parameter $\phi_j$ specifies the cost cut induced by each link of loan due to the collaboration between banks.
Hence, since the profit is increasing in the links between banks, it implies the local strategic  complmentarity. 
Also, since the price is decreasing in aggregate quantity of loans, there is the global strategic substitutability.
Under competition, each bank decides the quantity of loan to maximize its profit, so the first order condition gives
\begin{equation}\label{eq:network game:interbank loan}
    x_j = (\th -  c_{0,j}) - \sum_{i=1}^n x_i + \phi_j \sum_{i=1}^n x_i g_{ij} =  \phi_j \lp  \sum_{i=1}^n x_i w_{ij} + \e_j \rp
\end{equation}
where we let $w_{ij} = g_{ij} - 1$ and $\e_j = (\th - c_{0,j}) / \phi_j$.
In this example, the interaction functions are different, since the coefficients $\phi_j$ may be heterogeneous.

\subsubsection{Network with Cross-Holdings}\label{section:CrossHoldings}
\cite{elliott2014financial} consider a financial network with cross-holdings and study the cascade effect of financial failure, which has the same form as input-output analysis (\ref{eq:input-output}).
In their framework, banks own some share of the other banks by lending or investment, so banks' values depend on other banks' holding assets.
They show that the cascade effect depends on network interconnections, in the sense that integration and diversification lead to different non-monotonic effects.
They consider an economy with $n$ financial institutions or banks, indexed $j =1, \dots, n $.
Each organization holds a basket of primitive assets, indexed $h=1, \dots, m$, which could be some projects that create cash flows.
The share of asset $h$ that organization $j$ holds is denoted $b_{hj} \geq 0$.
The market price of the asset $h$ is denoted as $p_h$.

Organizations crossly hold some shares of the other organizations in the networks.
For all $i, j = 1, \dots, n$, let $d_{ij}$ be the debt that organization $i$ has to repay to $j$ or the amount of fund invested in organization $i$ by organization $j$.
Define $w_{ij} \coloneqq d_{ij}/x_i$.
Hence, organization $j$ owns $w_{ij} \in [0, 1)$ fraction of the values of organization $i$.
Assume that $w_{ii}=0$ for all $i$.
Denote the book value or equity of organization $j$ as $x_j$.
The book value $x_j$ is the total asset value that $j$ owns (i.e., the book total value of its primitive assets and its claims on other organizations.)
\begin{equation}\label{eq:cross-holding:equity}
    x_j = \sum_{h=1}^m p_h b_{hj} + \sum_{i=1}^n d_{ij} = \e_j + \sum_{i=1}^n x_i w_{ij} 
\end{equation}
where $\e_j = \sum_{h=1}^m p_h b_{hj}$.
Assume that external investors hold strictly positive shares of organization $i$, i.e., $1 - \sum_{j \in I} w_{ij} > 0$. 
Then, $W$ is non-singular and the equilibrium equity is given by $x = \e (I - W)^{-1}$.

\subsubsection{Financial Network}\label{section:Financial}
This section introduces the financial networks of interbank lending from \cite{eisenberg2001systemic} and \cite{Financial:cifuentes2005liquidity}.
It is used to study the contagion of default under the
conditions of proportional repayments of liabilities, limited liability, and absolute priority of debt over equity.
For instance, \cite{Financial:glasserman2015likely} bound the probability of default due to contagion when there is a bank that suffers the shock.

There are $n$ risk-neutral banks in the network as the previous section. 
Each bank $i$ has the nominal liability $\d_{ij}$ to bank $j$.
The total liability obligation of $i$ is $\bar p_i = \sum_{j} \d_{ij}$.
Define $w_{ij} = \d_{ij} / \bar p_i$ if $\bar p_i > 0$ and $w_{ij} =0$ otherwise.\footnote{
The sensitivity matrix $(w_{ij})$ is also called as the \emph{relative liability matrix} in Eisenberg-Noe model.}
Assume that $\sum_j w_{ij} = 1$ for all $i$, in the sense that there is no payment to agents outside the network.
All banks have the exogenous cash flow $\e_j \geq 0$, which can be interpreted as the net asset from outside the financial network.
Let $x_j$ be the clearing repayment for all $j=1, \dots, n$ in equilibrium.
The amount of total repayment received by $j$ from other banks is $\sum_i x_i w_{ij}$.
Suppose that in equilibrium all banks follow the conditions of limited liability, $x_j \leq \sum_i x_i w_{ij} + \e_j $, and absolute priority, either $x_j = \bar p_j$ or $x_j =\sum_i x_i w_{ij} + \e_j$.
The clearing payment $x_j$ in equilibrium solves
\begin{equation} \label{eq:EN model}
  x_j = \min\left\{\sum_i x_i w_{ij} + \e_j  , \bar p_j \right\}
\end{equation}
for all $j$.
Observe that the interaction functions are
 $$f_j(t) =  t \, \mathbbm{1}_{\{t < \bar p_j\}}(t) + \bar p_j \, \mathbbm{1}_{\{t\geq \bar p_j\}}(t)$$
for all $j$.
\cite{eisenberg2001systemic} show that the clearing payment is unique if the financial network is regular.\footnote{The \textit{risk orbit} of the bank $i$  is the set that $i$ has a directed path to all nodes in the set. The system is \emph{regular} if any risk orbit has at least one node $i$ with positive cash flow $e_i > 0$.}
\cite{Financial:glasserman2015likely}, instead, assume that for every bank $i$ either it has some debt outside the network $\sum_j w_{ij} < 1$, or there is a bank $t$ such that $i \ra \cdots \ra t$ and bank $t$ has external debt $\sum_j w_{tj} < 1$. \footnote{That is, $w_{it}^k  > 0$ for some $k \in \N$ and $\sum_j w_{tj} < 1$.}
Under this assumption, $W$ is convergent and then the clearing payment is also unique (Lemma \ref{condition:convergence} holds.)

\cite{acemoglu2015systemic} and \cite{acemoglu2015Endogenous} consider the Eisenberg-Noe network with senior liability and asset liquidation and show that the contagion of financial default depends on both the magnitude of shock and the network structure.
In their setting, the shock $\e_j$ could be negative. 
In detail, each bank can liquidate their asset to pay the debt. 
Let $\ell_j$ be the liquidation decision for all $j$.
Assume that banks can only recover $\zeta \in [0, 1]$ fraction of the value of a liquidated project.
The repayment decision is
$$x_j = \max \left\{ \min\left\{ \sum_i x_i w_{ij} + c_j + z_j - \nu + \zeta \ell_j, \, \bar{p}_j\right\}, 0 \right\}$$
where $c_j$ is the cash, $z_j$ is project return and, $\nu$ senior liability.
A bank's ability to fulfill its liability depends on its resource,  including the received repayments from the bank's debtors, its hoarding cash, the return of the invested project minus the senior liability, and the liquidated asset of the project.
Except for the rules of limited liability and absolute priority, banks repay nothing if the total cash flow is negative that $\sum_i x_i w_{ij} + c_j + z_j - \nu + \zeta \ell_j < 0$.
Moreover, each bank decide the amount of liquidation:
$$\ell_j = \max\left\{\min\left\{\frac{1}{\zeta} \lp\bar p_j - \sum_i x_i w_{ij} - e_j\rp, A\right\}, 0 \right\} $$
where $e_j = c_j + z_j - \nu$ and $A$ is the total value of the invested project.
A Bank can liquidate a fraction of its invested project, with value $A$ to meet the shortfall of liability $\bar p_j- \sum_i x_i w_{ij} - e_j$. 
\cite{acemoglu2015systemic} show that the payment in equilibrium satisfies:\footnote{See Lemma B2 of \cite{acemoglu2015systemic}}  
\begin{equation} \label{eq:generalised EN model}
  x_j = \max \left\{ \min\left\{\sum_i x_i w_{ij} + \e_j ,\,\, \bar p_j \right\}, 0   \right\}  
\end{equation}
where $\e_j= c_j + \xi_j - \nu + \zeta A $ could be negative.
In particular, the interaction functions are
 $$f_j(t) =  t \, \mathbbm{1}_{\{0 < t < \bar p_j\}}(t) + \bar p_j \, \mathbbm{1}_{\{t\geq \bar p_j\}}(t)$$
for all $j$. 
\cite{acemoglu2015systemic} show that the clearing payment is generically unique for a strongly connected network.\footnote{See Definition \ref{def:generic uniqueness:Lebesgue} for the definition of generic uniqueness.}
Their proof also shows that the clearing payment (\ref{eq:EN model}) of Eisenberg-Noe network is unique, given a strongly connected network.

\subsubsection{Financial Network with Bankruptcy Cost}\label{section:Financial:BankruptcyCost}
This subsection introduces another generalized Eisenberg-Noe model with bankruptcy cost in \cite{Financial:glasserman2015likely} and \cite{ financial:glasserman2016contagion}.\footnote{The bankruptcy costs includes auditing, accounting, and legal costs, and the losses associated with asset liquidation.}
Consider a relative liability matrix $(w_{ij})$ as previous subsection.
Suppose that when each bank $j=1, \dots, n$ defaults, its asset is further reduced by $$\a_j \lb \bar p_j - \lp \sum_i x_i w_{ij} + \e_j \rp \rb$$
up to a maximum reduction that the assets are entirely eliminated.
In other words, a large shortfall of liability generates a higher bankruptcy cost than a small shortfall.
Then, the clearing payment is 
\begin{equation} \label{eq:EN model:bankrupty cost}
\begin{split}
  x_j &= \min \left\{ \max\left\{\sum_i x_i w_{ij} + \e_j - \a_j \lb \bar p_j - \lp \sum_i x_i w_{ij} + \e_j \rp \rb , 0 \right\}, \bar p_j   \right\}  \\
 & =\min \left\{ \max\left\{ (1+\a_j)\lp \sum_i x_i w_{ij} + \e_j \rp - \a_j \bar p_j, \; 0 \right\}, \bar p_j   \right\}  
\end{split}
\end{equation}

Following \cite{Financial:glasserman2015likely} and \cite{financial:glasserman2016contagion}, when $\a_i \equiv \a$, the clearing payment is unique if $(1 + \a) \max_i \sum_{j} w_{ij} < 1$, which implies that $(1+\a) W$ is non-singular.
Hence, the interaction functions can be written as\footnote{For the last term, $\mathbbm{1}_{ \{(1+\a_j) t - \a_j \bar p_j \geq \bar p_j \}} (\bar p_j) = \mathbbm{1}_{\{ t  \geq \bar p_j\}}(\bar p_j)$.}
$$f_j (t) = ((1+\a_j)t - \a_j \bar p_j) \mathbbm{1}_{ \{0 \leq (1+\a_j) t - \a_j \bar p_j < \bar p_j \}} (t) + \bar p_j \mathbbm{1}_{\{ t  \geq \bar p_j\}}(t)$$ 
for all $j$, where $\mathbbm{1}_A (t)$ is the indicator function with $A \subset \R$.

Alternatively, \cite{financial:rogers2013failure} consider the following payment function with $\e_j \geq 0$:
\begin{equation*} 
x_j = 
    \begin{cases}
        \bar p_j  & \text{if }  \sum_i x_i w_{ij} + \e_j \geq \bar p_j \\
            \a \e_j + \beta \sum_i x_i w_{ij} & \text{otherwise.}
    \end{cases}
\end{equation*}
for all $j$, where $0 < \a, \b < 1$.
This payment function (\ref{eq:financial:rogers_veraart}) is discontinuous, since $\a, \b < 1$.
Under this payment function, the bankruptcy cost is $(1 - \a) \e_j + (1-\b) \sum_i x_i w_{ij}$.\footnote{The bankcuptcy cost is $\sum_i x_i w_{ij} + \e_j - x_j$ when the bank defaults.}

Given $\e_j$, let $q_j = \bar p_j + (\a / \b -1) \e_j$ and $e_j = \a \e_j / \b$.
We can rewrite the payment function 
\begin{equation} \label{eq:financial:rogers_veraart}
x_j = 
    \begin{cases}
        \bar p_j  & \text{if }  \sum_i x_i w_{ij} + e_j \geq q_j \\
        \b \lp\sum_i x_i w_{ij} + e_j \rp & \text{otherwise.}
    \end{cases}
\end{equation}
The interaction function is 
$$f_j(t) = \b t \, \mathbbm{1}_{\{t < q_j\}}(t) + \bar p_j \mathbbm{1}_{\{t \geq q_j \}}(t)$$
\cite{financial:rogers2013failure} show that the clearing vector in equilibrium exists, but may not be unique.

\subsubsection{Financial Network with Equity Insolvency and Illiquidity}\label{section:financial:Liu et al}
Following the Eisenberg-Noe model (\ref{eq:EN model}), \cite{Financial:liu2020interbank} consider a financial lending network that banks are exposed to lending and borrowing with different maturities.
They show that the U.S. banking network has diminished its system risk of contagion and illiquidity from 2011 to 2014.

Consider that for each bank $i$ the asset in its balance sheet equals to the sum of overnight lending, short-term lending, long-term lending, cash and cash equivalents $\e_i$ and other assets $OA_i$, while the liability consists of overnight borrowing, short-term borrowing, other liability $OL_i$ and equity $E_i$.
Each period $t$, bank $i$ has the obligation to repay some fraction of overnight, short-term and long-term liability $\d_{ij}$ to bank $j$.
The total liability obligation is $\bar p_i$ in the period $t$.
The relative liability matrix $(w_{ij})$ is defined as before that $w_{ij} = \d_{ij} / \bar p_i$ if $\bar p_i > 0$ and $w_{ij} =0$ otherwise, so that $(w_{ij})$ is non-negative.
Assume that $\sum_j w_{ij} \leq 1$ following \cite{Financial:liu2020interbank}.
Let $Q_{ij}$ be the remainder of all loan obligations that $i$ has to repay $j$, including overnight market, short-term and long-term loans, at the end of period. 
Let $x_i$ be the realized payment made at the end of the period.
Define the equity 
$$E_i = \sum_{h}x_h w_{hi}  + \e_i - \bar p_i + \lp \sum_{h} Q_{hi} - \sum_{j} Q_{ij} \rp + \lp OA_i -  OL_i\rp$$
Denote $B_i \coloneqq \sum_{h} Q_{hi} - \sum_{j} Q_{ij}  +  OA_i -  OL_i$ as the net remaining and other assets.
Each bank fails to repay in full if either it is illiquid due to insufficient cash and incoming payment, or it is insolvent that its equities are negative $E_i < 0$.
The payment in equilibrium satisfies:
\begin{equation}\label{eq:financial:Liu et al}
  x_i = \min\left\{ \lb \sum_{h} x_h w_{hi} + \e_i \rb^+ ,  \lb \sum_{h}x_h w_{hi} + \e_i + B_i \rb^+,  \;\bar p_i \right\}  
\end{equation}
where $[z]^+ \coloneqq \max\{z, 0 \}$ for $z \in \R$.
The interaction functions are 
$$f_i(t) = \min\left\{ \lb t \rb^+ ,  \lb t + B_i \rb^+,  \;\bar p_i \right\}$$ for all $i$.
\cite{Financial:liu2020interbank} show that the equilibrium payment exists.
We further show that it is generically unique when $W$ is stochastic and unique when $W$ is convergent (see Section \ref{sec:generic uniqueness}).

%% file: uniqueness.tex
\section{Existence and Uniqueness}\label{sec:generic uniqueness}
In this section, we first show that the contraction of both the interaction functions and the sensitivity matrix could lead to the existence and uniqueness of equilibrium, under heterogeneous interaction functions.
Next, we show that the  equilibrium is unique almost surely when the interaction functions are increasing and all non-expansive but bounded, the spectral radius of the sensitivity matrix is equal to one, and the shock is absolutely continuous.
We assume the absolute continuity of shock to preclude the case that multiple equilibria occur with non-zero probability.
More, our results apply to any network structure.\footnote{We do not assume the strong connection in the case of non-contracting conditions.}
All the proofs are presented in the Appendix.

Observing the example models in Section \ref{sec:NetworkModels}, there are generally two categories of assumptions.
In the first category, the interaction functions and sensitivity matrix exist contraction property.
For the second category, the interaction functions are non-expansive but bounded, and the sensitivity matrix is non-convergent (i.e., stochastic or $r(W)=1$.)
In detail, we have the following two assumptions.\footnote{
Recall that the function $f_i \colon \R \ra \R$ is bounded if there is $M>0$ such that $|f_i(t)| \leq M$ for all $t\in \R$.
The function $f_i$ is Lipschitz continuous with Lipschitz constant $\b_i$ if $|f_i(x) - f_i(y)| \leq \b_i |x - y|$ for all $x, y \in \R$.
Moreover, $f_i$ is non-expansive if $\b_i = 1$.}

\begin{assumption}[Eventually Contracting]\label{assumption:contracting}
$f=(f_i)$ and $W$ satisfy~
\begin{enumerate}[label={\upshape(\roman*)}]
    \item $f_i$ is Lipschitz continuous with Lipschitz constant $\b_i$ for all $i \in N$, and
    \item $r(|W| \diag(\b)) < 1$, where $\b = (\b_i)$.
\end{enumerate} 
\end{assumption}

\begin{assumption}[Non-contracting]  \label{assumption:non-expansive}
$f=(f_i)$ and $W$ satisfy~
\begin{enumerate}[label={\upshape(\roman*)}]
\item $f_i$ is increasing, non-expansive and bounded for all $i$, and
\item $W$ is non-negative and $r(W)=1$.
\end{enumerate} 
\end{assumption}

We summarize that the networks (\ref{eq:input-output}), (\ref{eq:production_clearing}), (\ref{eq:production:log output}), (\ref{eq:networkgame:best-reply}), (\ref{eq:networkgame:best-reply:general}), (\ref{eq:network game:interbank loan}),  (\ref{eq:cross-holding:equity}) and (\ref{eq:EN model:bankrupty cost}) 
satisfy Assumption \ref{assumption:contracting}, and the networks (\ref{eq:EN model}), (\ref{eq:generalised EN model}) and (\ref{eq:financial:Liu et al}) satisfy Assumption \ref{assumption:non-expansive}.

We say that the network is \emph{(eventually) contracting} if $f$ and $W$ satisfy Assumption \ref{assumption:contracting} and is \emph{non-contracting} if $f$ and $W$ satisfy Assumption \ref{assumption:non-expansive}.
Comparing the two assumptions, note that Assumption \ref{assumption:contracting} does not require the boundedness for the interaction functions, and the sensitivity matrix can be negative.
While, Assumption \ref{assumption:non-expansive} supposes that the interaction functions are increasing and bounded, and the sensitivity matrix is non-negative and its spectral radius is one.
In application, the spectral radius is one as long as the row/column sum is one.
As shown in the subsequent propositions, if a network is eventually contracting, the map $x \mapsto f(xW + \e)$ is a Banach contraction and hence the equilibrium is unique.
On the other hand, if a network is non-contracting, there exists an almost surely unique equilibrium.

One exceptional model which does not satisfy either of the above assumptions is the network (\ref{eq:financial:rogers_veraart}) due to discontinuity.
However, since its interaction functions are increasing and bounded as Assumption \ref{assumption:non-expansive}, it can be shown that the greatest equilibrium and the least equilibrium exist.
In general, any network with increasing and bounded interaction functions admit the greatest and the least equilibrium.

\begin{lemma}\label{lemma:existence}
If $f_i$ is increasing and bounded for all $i \in N$, then the greatest and the least equilibria exist.
\end{lemma}


We first consider the eventually contracting network and discuss the cases that $f$ and $W$ satisfy Assumption \ref{assumption:contracting} such that $r(|W| \diag(\b)) < 1$.
One special case is that the contracting condition can merely depend on the interaction functions.
Taking the production network (\ref{eq:production:log output}) for example, $W$ is non-negative and its row sum is one, and $\b_i \equiv (1-\a)$, the spectral radius condition is reduced to $r(|W| \diag(\b)) = (1-\a) r(W) = (1-\a) < 1$.
In this case, the Lipschitz contraction determines the uniqueness of equilibrium.

Alternatively, the contracting condition may depend only on the convergence of the sensitivity matrix.
Taking the financial network (\ref{eq:EN model}) for example, \cite{Financial:glasserman2015likely} assume that the interaction functions are non-expansive, and the sensitivity matrix is non-negative and weekly chained substochastic such that $r(W) < 1$.
Hence, since we have $r(|W| \diag(\b) ) = r(W)<1$, the uniqueness of equilibrium depends on the convergence of the sensitivity matrix. 
When the interaction functions are non-expansive, we can check whether the matrix $W$ is weakly chained substochastic and apply Lemma \ref{condition:convergence} to show the convergence of $W$.



We briefly discuss some intuition behind Lemma \ref{condition:convergence} and weekly chained convergence.
The convergence can be seen by considering a simple case: if $\sum_j w_{ij} \leq 1$ for all $i$, and for all $i \in N$  there is $t\in N$ such that $i \ra t $ in one step $(w_{it} > 0)$ and $\sum_j w_{tj} < 1$, then we can show that $\sum_{k} w^2_{ik} = \sum_k \sum_j w_{ij} w_{jk} = \sum_j w_{ij} \sum_k  w_{jk} =  \sum_{j \neq t} w_{ij} \sum_k  w_{jk} + w_{it} \sum_k w_{tk} < 1$ for all $i$.
Hence, the row sums of $W^2$ are all strictly less than one.
It implies that $\| W^2\|_\infty < 1$ so that $r(W)<1$ and $W$ is convergent.
In application, we see that the assumptions of \cite{Financial:glasserman2015likely} about the Eisenberg-Noe financial network (\ref{eq:EN model}) satisfy weakly chained stochastic condition.

We summarized the discussion by the following lemmas, which show that some network structure of $\graph W$ determines the convergence.
For instance, Assumption \ref{assumption:contracting} holds as long as the network structure is acyclic such that there exists no feedback effect.

\begin{lemma}\label{corollary:convergence condition}
If $f_i$ is non-expansive for all $i\in N$, and $W$ is non-negative and weakly chained substochastic, then Assumption \ref{assumption:contracting} holds.
\end{lemma}

\begin{lemma}\label{lemma:substochasitc_irreducible}
If $f_i$ is non-expansive for all $i\in N$, $W$ is non-negative, row (column)  substochastic and irreducible,  and there is $t \in N$ such that $\sum_{j} w_{tj} < 1$ ($\sum_{j} w_{jt} < 1$), then Assumption \ref{assumption:contracting} holds.
\end{lemma}

\begin{lemma}\label{corollary:acyclic}
If $f_i$ is Lipschitz continuous for all $i$, and $W \in \R^{n \times n}$ is such that $\graph{W}$ is acyclic, then Assumption \ref{assumption:contracting} holds.
\end{lemma}

Generally, unlike \cite{acemoglu2016network}, the contraction condition $r(|W| \diag{(\b)}) < 1$ depends on both interaction functions and the sensitivity matrix.
For instance, we may have 
\begin{equation*}
\begin{split}
    &    W = \begin{pmatrix}
        0 & 2 \\
        4/7 & 0 
    \end{pmatrix}, \qquad
    \diag(\b) = \begin{pmatrix}
        5/4 & 0 \\
        0 & 2/3 \\
    \end{pmatrix}, \quad
     W \diag(\b) = \begin{pmatrix}
        0 & 4/3 \\
        5/7 & 0
    \end{pmatrix}.
    \end{split}
\end{equation*}
In this case, we have $r(W) > 1$, $\b_1 > 1$ and $r(|W| \diag(\b)) < 1$.

The spectral radius condition considers the sensitivity matrix by taking the absolute values of all entries $|W|$.  
We can reduce this condition when $W$ is symmetric. For instance, if $W \in \R^{n\times n}$ is symmetric and $\b_i \equiv \phi$, then the condition is $r(|W|\diag{(\b)}) = \phi r(W) < 1$.
We show the existence and uniqueness of equilibrium for an eventually contracting network.

\begin{proposition}\label{proposition:contraction_uniqueness}
If Assumption \ref{assumption:contracting} holds , then the equilibrium exists and is unique for any  $\e \in \R^n$.
\end{proposition}

Hence, an eventually contracting network with $r(|W| \diag(\b))  < 1$ always has a unique equilibrium.
Proposition \ref{proposition:contraction_uniqueness} also implies that we can compute the unique equilibrium by iteration. 
Define the mapping $T: \R^n \ra \R^n$ as 
\begin{equation}\label{eq:operator Tx=f(xW + e)}
    T x \coloneqq f(xW + \e).
\end{equation}
We say that $T$ is \emph{globally stable} on $\R^n$ if $T$ has a unique fixed point $x^* \in \R^n$ and $T^k x \ra x^*$ as $k \ra \infty$ for any $x \in \R^n$.

Since the map $T$ is a Banach contraction following the proof of Proposition \ref{proposition:contraction_uniqueness}, we can compute the equilibrium by iteration is $x^* =\lim_{k\ra \infty} T^k x$ by any initial guess $x \in \R^n$, whence $T$ is globally stable. 
We summarize the result in the following corollary.
\begin{corollary}\label{corollary:globally stable}
Suppose that Assumption \ref{assumption:contracting} holds. Then, we have
    \begin{enumerate}[label={\upshape(\roman*)}] 
    
        \item $T$ is globally stable. 
        \item $\| T^{k+1} x - T^k x\| \leq \| (|W|\diag{(\b)})^k \| \|Tx - x\| $ for all $k \in \N$ and $x \in \R^n$.
    \end{enumerate}
\end{corollary}

Corollary \ref{corollary:globally stable} provides the speed of convergence.
Since $r(|W|\diag{(\b)}) < 1$, we can find $m \in \N$ such that $q = \| (|W|\diag{(\b)})^m \| < 1$.
Then, corollary \ref{corollary:globally stable} implies that the $\| T^{k m + 1} x - T^{km}x \| \leq \| (|W|\diag{(\b)})^{km} \| \|Tx - x\| \leq q^k \| Tx - x\|$ for all $x \in \R^n$ and $k \in \N$.

Further, some networks may have more strict conditions on interaction functions.
Except for the Lipschitz continuity, the interaction functions of (\ref{eq:input-output}) - (\ref{eq:cross-holding:equity}) are linear such that $f_i(a) - f_i(b) = \b_i (a - b)$ for some $\b_i > 0$, for all $a, b \in \R$ and for all $i \in N$.
In this case, we can reduce the condition to $r(W \diag{(\b)} ) < 1$, by the similar proof as Proposition \ref{proposition:contraction_uniqueness}.\footnote{Note that the entry of $W$ could be negative as Assumption \ref{assumption:contracting}.}

\begin{corollary}\label{corollary: r(W diag(b))<1}
If for all $i\in N$ there is $\b_i > 0$ such that $f_i(a) - f_i(b) = \b_i (a - b)$ for all $a, b \in \R$, then for $W\in \R^{n \times n}$ the condition $ r(W \diag{(\b)} ) < 1$ implies the uniqueness of equilibrium for any $\e \in \R^n$.
\end{corollary}

The rest of this section investigates the non-contracting networks, which have the non-expansive interaction functions and the sensitivity matrix of spectral radius one.
In particular, we are interested in the networks like equations (\ref{eq:EN model}), (\ref{eq:generalised EN model}) and (\ref{eq:financial:Liu et al}) which follow Assumption \ref{assumption:non-expansive}.
In these networks, since the sensitivity matrices are stochastic, their spectral radii are equal to one.

We observe that when the networks are non-contracting, it is possible that they admit multiple equilibria with non-negative probability, as the following example.

\begin{example}\label{example:counter}
This example demonstrates that when the interaction functions are non-expansive, and the sensitivity matrix is stochastic such that $r(W) = 1$, there may be multiple equilibria.
Furthermore, the multiple equilibria may occur with non-zero probability if the shock variables are discrete.

Suppose that there are two agents in the economy, $n=2$.
Let $w_{12} = w_{21} = 1$ and $w_{11} = w_{22} = 0$ satisfy $\sum_h w_{hi} = 1$ for $i=1, 2$ (see Figure \ref{fig:counterexample}).
Consider the i.i.d. shock $\e_i \in \{1, -1\}$  for all $i$ with the equal positive probabilities, $\Prob(\e_i=1 )= \Prob(\e_i= -1 )=1/2$.
Then, the shocks $\e_i$ have mean zero and constant variance so that the conditions in \cite{acemoglu2016network} are satisfied. 
Also, consider the interaction function as $f_i \equiv g$ for all $i$ and 
\begin{equation} \label{eq:identity map}
 g(z) = -M \mathbbm{1}_{\{z < -M\}}(z) + z \mathbbm{1}_{\{-M \leq z \leq M\}}(z) + M \mathbbm{1}_{\{z> M\}}(z),
\end{equation}
where we set $M < \infty$.
Then, $g$ is a bounded and non-expansive identity mapping  (i.e., $g(z) = z$ if $|z| < M$ and $|g(z)| \leq M$ for all $z\in \R$.)
\begin{figure}[tb!]
    \centering
    \begin{subfigure}[tb]{0.4\linewidth}
        \centering
        \begin{tikzpicture}[baseline, auto, node distance={20mm}, thick, main/.style = {draw, circle, scale=0.8},] 
        \node[draw, circle, label={$\e_1=1$}] (1) at (0, 0) {$1$};
        \node[draw, circle, label={$\e_2=-1$}] (2) at (4, 0) {$2$};
        \path [->,bend left=10, above] (1) edge node {$w_{12}=1$} (2);
        \path [->,bend left=10, below] (2) edge node {$w_{21}=1$} (1);
        \end{tikzpicture}
    \end{subfigure}
\Large
\begin{subfigure}[tb]{0.4\linewidth}
    \centering
    \begin{tikzpicture}[baseline, scale=0.45, every node/.style={scale=0.6}]
        \draw [->] (-5,0) -- (5,0) node[below right] {$x$}; 
        \draw [->] (0,-3) -- (0,3) node[left]{$y$};
        \draw  [thick] (-5, -2) -- (-2, -2) -- (2, 2);
        \draw [thick] (2,2) -- (5, 2) node [above] {$g(z)$};
        \draw [dashed] (2, 2) -- (0, 2) node [left] {$2$};
        \draw [dashed] (-2, -2) -- (0, -2) node [right] {$-2$};
        \end{tikzpicture}
    \end{subfigure}    
    \caption{}
    \label{fig:counterexample}
\end{figure}

Suppose that $x=(x_i)$ satisfy $x_1 = g(x_2 + \e_1)$ and $x_2 = g(x_1 + \e_1)$.
Since $g$ is bounded, we have $-M \leq x_1, x_2 \leq M$.
If the realization is $\e = (\e_1, \e_2) = (1, -1)$, then the system $x = f(xW + \e)$ gives
\[
\begin{split}
x_1 = x_2 + 1   \\
x_2 = x_1 - 1
\end{split}
\]
where we assume $-M \leq x_2 + 1, x_1 -1 \leq M$ and then check the solutions.
Thus, the solutions are $x_1 = y + 1$ and $x_2 = y$ with $ -M\leq y \leq M-1$, so there are multiple equilibria if $\e = (1 , -1)$.
Similarly, there are multiple equilibria if $\e = (-1, 1)$.
Hence, the set of realization of the shocks generating multiple solutions is $S=\{\e \in \R^2: \e=(1, -1) \text{ or } (-1, 1) \}$, which is measure zero $\lambda(S) = 0$.
However, the probability of $\e$ that fails the uniqueness is $\Prob(\e \in S )= 1/2$, which is non-zero.
\qed
\end{example}

To avoid the confusion as Example \ref{example:counter} that multiple equilibria  occur with non-zero probability, we suppose that the idiosyncratic shocks are \emph{absolutely continuous}.\footnote{Recall that the random variable $\e_i$ is \emph{absolutely continuous}
if there is a Lebesgue integrable function $g$ such that $
\Prob (\e_i \in A) = \int_A g(x) \de \lambda$ for all Borel sets $A$ and for all $i \in N$.}
When the shocks are absolutely continuous, the events with measure zero are also occurring with zero probability.
Hence, the absolute continuity precludes the case of Example \ref{example:counter}.
The subsequent proposition shows the network has almost surely unique equilibrium when Assumption \ref{assumption:non-expansive} holds, and the shock is absolutely continuous.

\begin{proposition} \label{proposition:generic uniqueness and uniqueness}
If Assumption \ref{assumption:non-expansive} holds, and the shock variables $(\e_i)$ are absolutely continuous,
then the equilibrium exists and is unique almost surely.
\end{proposition}

By the Tarski's Fixed Point Theorem, since $f_i$ is continuous for all $i$, we can compute the equilibrium by iteration.
In particular, let $\ell_j$ and $u_j$ be the lower bound and upper bound of the interaction function for all $j$, if the network satisfies Assumption \ref{assumption:non-expansive}.
Then, denoting $u \coloneqq (u_j)$ and $\ell \coloneqq (\ell_j)$, the largest equilibrium is $x^* = \lim_{m \ra \infty} T^m u$, where $T: [\ell, u] \ra [\ell, u]$ is defined as (\ref{eq:operator Tx=f(xW + e)}).
If the shocks are absolutely continuous, then Proposition \ref{proposition:generic uniqueness and uniqueness} shows that $x^*$ is the almost surely unique equilibrium.
Hence, the largest equilibrium computed by the iteration from upper bound is the unique equilibrium.

\begin{corollary} \label{corollary:iteration from above}
If the assumptions of Proposition \ref{proposition:generic uniqueness and uniqueness} hold, then the almost surely unique equilibrium is $\lim_{n \ra \infty}T^n u.$
\end{corollary}

Unlike the theorem in \cite{acemoglu2016network}, Proposition \ref{proposition:generic uniqueness and uniqueness} allows the sensitivity matrix to be either row or column stochastic and not necessarily strongly connected.
The sensitivity matrix can also be non-stochastic as long as its spectral radius is one (see Example \ref{example:comparative statics}).
In the following example, we apply Proposition \ref{proposition:generic uniqueness and uniqueness} to show that the network (\ref{eq:financial:Liu et al}) has almost surely unique equilibrium.

\begin{example} \label{example:proposition 4.2}
Consider the financial network (\ref{eq:financial:Liu et al}), we have $0 \leq W \leq 1$ and for agent $j$
\begin{align*}
  x_j & = \min\left\{ \lb \sum_{i} x_i w_{ij} + \e_j \rb^+ ,  \lb \sum_{i}x_i w_{ij} + \e_j + B_j \rb^+,  \;\bar p_j \right\}  \\
  &= 
  \begin{cases}
  \min\left\{ \lb \sum_{i} x_i w_{ij} + \e_j \rb^+,  \;\bar p_j \right\} & \text{ if $B_j \geq 0$, }\\
  \min\left\{  \lb \sum_{i}x_i w_{ij} + \e_j + B_j \rb^+,  \;\bar p_j \right\} & \text{ otherwise.}
  \end{cases}
\end{align*}
Then, the interaction function is
\begin{equation*}
    f_j(z) = 
    \begin{cases}
    z \mathbbm{1}_{\{0\leq z < \bar p_j\}}(z) + \bar p_j \mathbbm{1}_{\{z\geq \bar p_j\}}(z)  & \text{ if $B_j \geq 0$,}\\
    (z + B_j) \mathbbm{1}_{\{0\leq z + B_j < \bar p_j\}}(z) + \bar p_j \mathbbm{1}_{\{z + B_j\geq \bar p_j\}}(z)  & \text{ otherwise.}
    \end{cases}
\end{equation*}
Clearly, $f_j$ is increasing and bounded for all $j$.
By figure \ref{fig:counterexample}, we know that the interaction function is non-expansive for either $B_j \geq 0$ or $B_j < 0$.
When the shock is absolutely continuous, Proposition \ref{proposition:generic uniqueness and uniqueness} shows that the clearing payment is unique almost surely when $W$ is stochastic.
On the other side, if $W$ satisfies the conditions in \ref{condition:convergence} or $r(W) < 1$, the clearing payment is unique by Proposition \ref{proposition:contraction_uniqueness}.
\qed
\end{example}

The proof of Proposition \ref{proposition:generic uniqueness and uniqueness} also implies that the network structure affects the existence of multiplicity.
We know that from Lemma \ref{corollary:acyclic} that if the network is acyclic, then the equilibrium must be unique.
Conversely, if there exist multiple equilibria, there must be some strongly connected subgraph.
The proof of Proposition \ref{proposition:generic uniqueness and uniqueness} further implies that all agents who have multiple equilibria must be in or accessible from some strongly connected subgraph, where all agents in this subgraph admit multiple equilibria.\footnote{
To prevent confusion, although we assume absolute continuity of shock in Proposition \ref{proposition:generic uniqueness and uniqueness}, we can specify the realized shocks or relax the absolute continuity so that there exists multiplicity, if we want.}

\begin{corollary}\label{corollary:multiplicity subgraph}
Let $(f, W)$ be such that Assumption \ref{assumption:non-expansive} holds.
Given the shocks, if there are multiple equilibria, then any agents having multiple equilibria must be accessible from some agents in a strongly connected subgraph $S$, which admits multiple equilibria.
\end{corollary}

In the rest part of this section, we apply the proof of Propoisition \ref{proposition:generic uniqueness and uniqueness} to show the uniqueness of \cite{eisenberg2001systemic} model, which is also shown in \cite{stachurski2022systemic}. 
The proof of Proposition \ref{proposition:generic uniqueness and uniqueness} implies that the set of shocks that admit the multiple equilibria is Lebesgue measure zero.
If we can further preclude the realization of such shocks, we can guarantee the uniqueness of equilibrium.
Eisenberg-Noe model (\ref{eq:EN model}) of clearing payment is a special case that it precludes the possibility of such shocks.
It can be shown that the multiple equilibria occur only if $\e \bm 1^\top = 0$ in Eisenberg-Noe network (\ref{eq:EN model}). 
Recall that the cash flows are non-negative $\e \geq 0$ in Eisenberg-Noe model.
Hence, if we further assume that there exists some $i \in N$ such that $\e_i > 0$, the equilibrium is always unique.

\begin{corollary}\label{corollary:EN uniqueness}
Let $f$, $W$ and $\e$ follow Eisenberg-Noe model (\ref{eq:EN model}).
That is, $f_j(t) =  t \, \mathbbm{1}_{\{t < \bar p_j\}}(t) + \bar p_j \, \mathbbm{1}_{\{t\geq \bar p_j\}}(t)$ for some $\bar p_j > 0$ for all $j$, $W$ is non-negative and row stochastic, and $\e \geq 0$.
Then, if $\e > 0$, the equilibrium is unique.
\end{corollary}

Since by convention we can set the equilibrium to be zero for all agents if $\e=0$ in Eisenberg-Noe model, Corollary \ref{corollary:EN uniqueness} implies that the equilibrium is unique for any $\e$. 
Therefore, the uniqueness of clearing payment holds without assuming that the network is regular as \cite{eisenberg2001systemic}.\footnote{\cite{stachurski2022systemic} shows the same result beautifully by Du's theorem.}
\cite{financial:staum2016systemic} assume that every bank has strictly positive external asset for the uniqueness of clearing payment.
\cite{financial:amini2016uniqueness} also assumes that either all banks hold external assets or the total of external assets is nonzero so that the conditions of in \cite{eisenberg2001systemic} are satisfied.\footnote{\cite{financial:amini2016uniqueness} studies the equilibrium of Eisenberg-Noe interbank network with asset liquidation, which affects the equilibrium price for the illiquid asset.}
Corollary \ref{corollary:EN uniqueness} implies that we only need one bank with positive external asset to have unique clearing payment.

\section{Comparative Statics, Tightness and Boundedness}\label{sec:tightness and boundedness}
In this section, we first study the comparative statics of how the increase in interaction functions, sensitivity matrix and shocks affect the equilibrium.
We further investigate the tightness of condition of Assumption \ref{assumption:contracting} and the requirement of boundedness in Assumption \ref{assumption:non-expansive}. We discuss an example to show that the spectral radius condition of Assumption \ref{assumption:contracting} could be also a necessary condition.
We then argue the boundedness condition is essential in Assumption \ref{assumption:non-expansive}.
In Section~\ref{sec:algorithm}, we consider an algorithm to compute the equilibrium when the interaction functions are bounded identity maps as financial network (\ref{eq:generalised EN model}).

\subsection{Comparative Statics}\label{sec:comparative statics}
This section presents some simple comparative statics between two networks.
We show that the equilibrium is increasing in the shock, $\e$, holding all other things constant.
Moreover, \textit{ceteris paribus}, when a network has a dominant interaction functions, it has a greater equilibrium. 
We also see that the rise in strength of interactions also increases the equilibrium.
These results are valid under both Assumption \ref{assumption:contracting} and Assumption \ref{assumption:non-expansive}.
In detail, we have the following two lemmas.

\begin{lemma}\label{lemma:comparative:contracting}
Let $(f, W, \e)$ and $(f', W', \e')$ be two networks satisfying Assumption \ref{assumption:contracting}, and denote their corresponding equilibrium as $\hat x$ and $\hat x'$, respectively.
If $f_i$ and $f'_i$ are increasing functions for all $i \in N$, 
$f_i(t) \leq f'_i(t)$ for all $t \in \R$ and all $i$, $W \leq W'$, and $\e \leq \e'$, then $\hat x \leq \hat x'$.
\end{lemma}

\begin{lemma}\label{lemma:comparative:non-contracting}
Let $(f, W, \e)$ and $(f', W', \e')$ be two networks satisfying Assumption \ref{assumption:non-expansive}
such that they have unique equilibrium, denoted by $\hat x$ and $\hat x'$, respectively.
Suppose that for all $i$ we have $f_i(t) \leq u_i$ and $f'_i(t) \leq u'_i$ for all $t\in \R$ such that $u_i \leq u_i'$.
If $f_i(t) \leq f'_i(t)$ for all $t \in \R$ and all $i$, $W \leq W'$, and $\e \leq \e'$, then $\hat x \leq \hat x'$.
\end{lemma}
These two lemmas imply that the equilibrium is increasing in the shock, the sensitivity matrix, or the interaction functions. 
When we add edges or increase the weight of edges in a network, since the interaction functions are increasing, the strengthened interconnections lead to greater equilibrium. 

About Lemma \ref{lemma:comparative:non-contracting}, note that it is possible that for two sensitivity matrices $W, W' \in \R^{n \times n}$ we have $r(W)=r(W')=1$ while $W \leq W'$.
For example, the sensitivity matrices in Example \ref{example:comparative statics} both have spectral radius one.
However, if we restrict the sensitivity matrix to a stochastic matrix, then we should fix $W=W'$ in Lemma \ref{lemma:comparative:non-contracting}, otherwise the row or column sum is not one anymore.

\begin{example}\label{example:comparative statics}
Consider the following specifications for $f, W$ and $\e$ as
\begin{align*}
    &f_i^a (t) \coloneqq \min\{ \max\{t, 0 \}, 2\},  \quad \forall i \qquad \e^a \coloneqq (0.2, -0.6, -0.2, 0.2)\\
    & f_i^a (t) \coloneqq \min\{ \max\{t, 0.1 \}, 2\}, \quad \forall i  \qquad \e^b \coloneqq (0.2, 0, -0.2, 0.2),
\end{align*}
and 
\begin{equation}\label{example:r(W)=1 matrix}
    W^a \coloneqq \begin{pmatrix}
    0 & 2 & 0 & 0\\
    0.5 & 0 & 0.5 & 0 \\
    0 & 0 & 0 & 0.8 \\
    0 & 0 & 0.8 & 0 \\
    \end{pmatrix}
    \qquad
    W^b \coloneqq \begin{pmatrix}
    0 & 2 & 0.1 & 0.8\\
    0.5 & 0 & 0.8 & 0.1\\
    0 & 0 & 0 & 0.9 \\
    0 & 0 & 0.9 & 0 \\
    \end{pmatrix}.
\end{equation}
The equilibrium for $(f^a, W^a, \e^a)$ is $(0.2, 0, 0, 0.2)$, the equilibrium for $(f^b, W^a, \e^a)$ is 
$(0.25, 0.1, 0.1, 0.28)$, the equilibrium for $(f^a, W^b, \e^a)$ is $(0.2, 0, 0.7579, 1.0421)$, the equilibrium for $(f^a, W^a, \e^b) $ is $(1.2, 2, 2, 1.8)$, and the equilibrium for $(f^b, W^b, \e^b)$ is $(1.2, 2, 2, 2)$.
Last, for the network with both $W^a$ and $W^b$, the multiplicity exists if $\e \in \R^4$ satisfies $2 \e_1 + \e_2 = 0$.
This condition does not hold almost surely if the shock is absolutely continuous. \qed
\end{example}

\subsection{Tightness of Condition}
In some cases, the spectral radius condition in Assumption \ref{assumption:contracting}, $r(|W|\diag(\b)
) < 1$ can be the necessary condition for the existence and uniqueness of equilibrium.
For example, consider the system with two agents such that the interaction function is $f_i(t) = \sqrt{t^2}$ for all $t \in \R$ for all $i = 1, 2$, and the realization of shocks are positive $\e_1, \e_2 > 0$. 
Assume that agents' states do not affect themselves and have positive influence on each others, such that $w_{11}= w_{22}=0$ and $w_{12}, w_{21} > 0$: 
\begin{equation*}
W = c
\begin{pmatrix}
0 & 1 \\
\lambda & 0
\end{pmatrix}
\end{equation*}
where $\l, c > 0$. 
Then, the interaction functions are Lipschitz continuous with a Lipschitz constant $\b_i = 1$ for $i=1, 2$.
Further, we can see that the eigenvalues of $W$ are $\pm c \sqrt{\l}$, whence the spectral radius is less than one, $r(W) < 1$, if and only if $c^2\l < 1$.
Now, the equilibrium follows 
\begin{align*}
&x_1 = \sqrt{(c \l x_2 + \e_1)^2} \\
& x_2 = \sqrt{(c x_1 + \e_2)^2}.
\end{align*}
for $x_1, x_2 \in [0, \infty)$.
Since we can show that this system has the unique solution if and only if $c^2 \l < 1$, the condition $r(W) < 1$ is a necessary condition for the existence and uniqueness of equilibrium.

As another example, considering the linear system $x = xW + \e$, the equilibrium could exist and be unique when $r(W) > 1$, but $T\colon x \mapsto f(xW + \e)$ is not globally stable.
In some applications, we want the equilibrium to be non-negative or positive.
For instance, the equilibrium in the production networks (\ref{eq:input-output}) and (\ref{eq:production_clearing}) should be non-negative.
Since $(I-W)^{-1} \geq 0$ if and only if $r(W) < 1$, Assumption \ref{assumption:contracting} is sufficient and necessary for non-negative equilibrium.\footnote{See Lemma \ref{lemma:non-negative I-W inverse}.}

In some other cases, the spectral radius condition is not a necessary condition for the existence.
The equilibrium may still exist when $r(|W|\diag(\beta)) \geq 1$.
However, the system is not globally stable anymore when the spectral radius is greater than one.
Also, the uniqueness of equilibrium could fail.
For example, suppose that the equilibrium satisfies
\begin{equation*}
    (x_1, x_2) = \min \left\{ \max\left\{ \left[(x_1, x_2) \begin{pmatrix}
    0 & 2 \\ 3 & 0
    \end{pmatrix}  + (\e_1, \e_2)\right],  (0, 0)\right\}    , (5, 5)\right\}.
\end{equation*}
The multiplicity and global stability depend on the shocks. 
For example, if we have $(\e_1, \e_1) = (-6, 2)$, then there are multiple equilibria $x^* = (0, 2)$ or $(5, 5)$.
More, it is not globally stable, since some iteration may not converge, say, the iteration starting from $(0, 5)$.

\subsection{Linear System and Boundedness Condition} \label{sec:linear system}
This part attempts to argue that when the interaction function is non-expansive and the sensitivity matrix is not convergent, the boundedness is essential for existence of the equilibrium.
That is, the boundedness condition in Assumption \ref{assumption:non-expansive} cannot be precluded.
We use a linear system to illustrate this concept.
In particular, if the interaction system is an identity mapping, and the sensitivity matrix has spectral radius one $r(W)=1$, then the equilibrium does not exist with probability one.

Consider a linear system: 
\begin{equation} \label{eq:linear system}
    x = x W + \e
\end{equation}
where $x$ and $\e$ are vectors in $\in \R^n$ and $W \in \R^n \times \R^n$ is a non-negative matrix.
By Proposition \ref{proposition:generic uniqueness and uniqueness}, we know that the linear system has a unique solution $x$ if the spectral radius of $W$ is less than one.
For instance, the study of input-output analysis assumes that the producers have positive value-added and then that $\sum_j w_{ij} < 1$ for all $i$ \citep{antras2012measuring}.
Note that (\ref{eq:linear system}) may not have the solution if scalar one is the eigenvalue of $W$, since $(I-W)$ is not invertible.
If one is not the eigenvalue of $W$, then the solution always exists and is unique.
Therefore, when the sensitivity matrix is stochastic, the solution does not exist almost surely.
Similar to almost sure uniqueness, we define almost sure non-existence as follows.

\begin{definition}
Let $E$ denotes the set of shocks that the solution exists (i.e., $E \coloneqq \{\e \in \R^n: \text{Equation (\ref{eq:linear system}) has a solution} \}$.)
We say that the solution \textit{does not exist almost surely} if $\Prob(\e \in E ) = 0$.
\end{definition}


\begin{lemma}
\label{lemma:linear:existence}
If $W$ is non-negative and $r(W)=1$, and the shocks $(\e_i)$ are i.i.d. and absolutely continuous, then the solution of linear system (\ref{eq:linear system}) does not exist almost surely.
\end{lemma}

Comparing Proposition \ref{proposition:generic uniqueness and uniqueness} and Lemma \ref{lemma:linear:existence}, we see that boundedness plays a key role to guarantee the existence of equilibrium.
Boundedness also helps to pin down the uniqueness of equilibrium.
We provide some intuition as below.
Consider again the interaction function of bounded identity map (\ref{eq:identity map}) and a row stochastic and irreducible matrix $W$.
From the proof of Lemma \ref{lemma:irreducible, generically unique}, we see that the multiple equilibria $x$ must satisfy $-M \bm{1} \leq x=xW + \e \leq M \bm{1}$.
Since $W$ is stochastic, $\bm{1}$ is the right eigenvector, whence we have $ \e \bm{1}^\top =(x - xW) \bm{1}^\top = 0$ if there exists multiplicity.
To notice how boundedness pins down the unique solution, suppose on the contrary that $\e \bm{1}^\top \neq 0$.
Thus, we have $xW + \e \nleq M \bm{1}$ or $-M \bm{1} \nleq xW + \e$.
Without loss of generality, assume that node $i$ is such that $(xW + \e)_i > M$, whence $i$'s equilibrium state equals $M$.
Remove such node $i$ from the graph and consider the shock $\e_s + x_i w_{is} $ for all remaining agents $s \neq i$. 
Let $\overline{W}$ be the submatrix of $W$ by  removing the column $i$ and row $i$ from $W$.
Using Lemma \ref{lemma:submatrix r(W)<1}, we have $r(\overline{W}) < 1$.
Therefore, the remaining network with vertices $N\setminus\{i\}$ has a unique solution by Proposition \ref{proposition:contraction_uniqueness}.
We see that when $\e \bm 1^\top \neq 0$ in this example, the boundedness pins down the unique solution.

\subsection{Algorithm for Bounded Identity Map}\label{sec:algorithm}
In this section, we provide an algorithm to compute the equilibrium under the bounded identity maps of interaction functions.
We show that the algorithm converges in at most $n 2^{n-1}$ iterations.
From the discussion in Section \ref{sec:linear system}, we know that the bounded condition is the key for the uniqueness of equilibrium.
It implies that we can compute the equilibrium by assuming that there exists an agent whose state is always equal to the upper or lower bound of her sensitivity function.
In detail, consider the interaction functions
\begin{equation}\label{eq:bounded identity map}
f_j (t) = \min \left\{ \max \left\{t, \ell_j\right\}, u_j \right\}
\end{equation}   
for all $j$, where $u_i > \ell_j$ for all $j \in \N$.
The bounded identity maps are the generalized Eisenberg-Noe model (\ref{eq:generalised EN model}).
Assume that the sensitivity matrix $W$ is row or column stochastic as the financial network.
Following the discussion in Section \ref{sec:linear system}, we conclude the lemma below.
Denote $u = (u_i)$ and $\ell=(\ell_i)$.

\begin{lemma} \label{lemma:unique solution at boundedness}
Let $u, \ell$ be such that $u \gg \ell$, $f$ be defined as (\ref{eq:bounded identity map}), and $W \geq 0 $ be row/column stochastic. 
Given $\e$, if the equilibrium $x^*$ is unique, then there is $j \in N$ such that either $x^*_j =  u_j$ or $x^*_j  = \ell_j$.
\end{lemma}

In fact, for any strongly connected subgraph $G_s \subset \graph W$, we can show that there exists an agent $j$ in subgraph $G_s$ such that $x_j^* = u_j$ or $x_j^* = \ell_j$.
Lemma \ref{lemma:unique solution at boundedness} implies that every agent's equilibrium state is either at the upper bound, lower bound, or in between. 
Since there are $n$ agents in the network, we have $3^n$ possibilities. 
Since we can pin down the equilibrium for those agents at the boundedness, we only need to decide the equilibrium for the rest agents.
Using this idea, we are able to design an algorithm that converges to the equilibrium in at most $3^n$ iterations.
Algorithm \ref{algorithm} applies this concept to search the equilibrium in finite iterations.

To illustrate the algorithm, given the states $x$, define the sets $A(x)$ and $B(x)$ as 
\begin{equation}\label{eq:sets of agents at boundedness}
    \begin{split}
    A(x) \coloneqq \{j \in N \colon \sum_{i} x_{i} w_{ij} + \e_j \geq u_j \} ,\\
    B(x) \coloneqq \{j \in N \colon \sum_{i} x_{i} w_{ij} + \e_j \leq \ell_j  \}.
\end{split}
\end{equation}
In words, $A(x)$ ($B(x)$) is the set of agents that their equilibria are greater (less) than or equal to the upper (lower) bounds under states $x$.
Also, define the diagonal matrix $\Lambda^{D}$ for $D \subset N$ as 
\begin{equation} \label{eq:diagonal:set at bound}
      \Lambda^D_{ij} \coloneqq \begin{cases}
    1 & i = j \text{ and } i \in D, \\
    0 & \text{ otherwise,}
    \end{cases} .
\end{equation}
Hence, $\Lambda^{A(x)}$ and $\Lambda^{B(x)}$ indicate which agent is at the upper and lower bound, respectively.

In the financial network, $N \setminus A(x)$ is the set of banks which default under the clearing payments $x$, while $N \setminus B(x)$ is the set of banks which are able to make some payments under the clearing payment $x$.
Inspired by \cite{eisenberg2001systemic}, Algorithm \ref{algorithm} returns the equilibrium $x^{(t)}$ given the interaction function (\ref{eq:bounded identity map}).
The inner for-loop (step 11-21) of Algorithm \ref{algorithm} is the fictitious default iteration introduced by \cite{eisenberg2001systemic}.
If we have $B(\ell) = \emptyset$, then Algorithm \ref{algorithm} is simply the fictitious default algorithm in \cite{eisenberg2001systemic}, which converges in at most $n$ iterations.
The outer for-loop (step 5-7) searches the set of agents whose equilibria are at the lower bounds, $B(x^*)$, from the potential candidates in the power set of $B(\ell)$, $\mathcal{P}(B(\ell))$. 
At some $t \in \N$, when the guess is correct: $P_i = B(x^*)$ for $P_i \in \mathcal{P}(B(\ell))$, we iterate the solutions from $\hat{u} \coloneqq u (I - \Lambda^{P_i}) + \ell \Lambda^{P_i}$,
and set $A_{t-1} = A(\hat{u}) \cap (N\setminus B(x^*))$.
Next, step $12$ is equivalent to set $x^{(t)}$ as 
\begin{equation*}
    x^{(t)}_j = 
    \begin{cases}
    u_j & \forall j \in A_{t-1} \\
    \ell_j & \forall j \in B^*\\
     \sum_{i \in A_{t-1}} u_i w_{ij} + \sum_{i \in B^*} \ell_i w_{ij} + \sum_{i \in N\setminus  (A_{t-1}\cup B^*)} x^{(t)}_{i} w_{ij} + \e_j & \text{otherwise.}
    \end{cases} 
\end{equation*}
Note that since $A_{t-1} \supset A(x^*)$, the matrix $I - (I -\Lambda^{A_{t-1}} -  \Lambda^{P_i}) W (I -\Lambda^{A_{t-1}} -  \Lambda^{P_i})$ in step $12$ is non-singular so that $x^{(t)}$ is unique:\footnote{We can use the iterative method to approximate the solution.}
\begin{multline*}
    x^{(t)}  = [((u \Lambda^{A_{t-1}} + \ell \Lambda^{P_i} ) W + \e )(I -\Lambda^{A_{t-1}} -  \Lambda^{P_i})  \\
    + u \Lambda^{A_{t-1}} + \ell \Lambda^{P_i} ] [I - (I -\Lambda^{A_{t-1}} -  \Lambda^{P_i}) W (I -\Lambda^{A_{t-1}} -  \Lambda^{P_i}) ]^{-1}.
\end{multline*}
The solution $x^{(t)}$ is the equilibrium when $x_i = u_i$ for $i \in A_{t-1}$ and $x_i = \ell_i$ for $i \in B^*$.
In the next step, we check whether $A_t = A(x^{(t)})$ equals $A_{t-1}$ or not.
If they are equal, the algorithm terminates and returns $x^{t}$ as the equilibrium; otherwise, we repeat the step $12$ of Algorithm \ref{algorithm} to get $x^{(t+1)}$ with the updated set $A_t$.
Conversely, if the guess from the power set of $B(\ell)$ is not correct, then it may be the case that it raises a singular matrix error in step $12$.
In this case, we skip it and try another guess from $\mathcal{P}(B(\ell))$.
Therefore, the convergence time depends on how many agents in $B(\ell)$.
In general, the next lemma shows that Algorithm \ref{algorithm} converges in at most $n 2^{n-1}$ iterations.\footnote{In some cases, the equilibrium $x^*$ has the features that $x^* \gg \ell$ or $x^* \ll u$.
If this is the case, then we can save time by implement the Eisenberg-Noe iteration and the reverse Eisenberg-Noe iteration firs.
}

\begin{lemma}\label{lemma:algorithm n2^n}
Let $f$ follow (\ref{eq:bounded identity map}), $W \geq 0$ be column/row stochastic, and $\e$ be such that the equilibrium is unique.
Algorithm \ref{algorithm} returns the equilibrium $x^{(t)}$ in at most $n 2^{n-1}$ iterations.
\end{lemma}

\begin{algorithm}[!htb]
\SetAlgoNoEnd%
\SetKwProg{try}{try}{}{}
\SetKwProg{except}{except}{}{}
\caption{Compute equilibrium given the interaction functions (\ref{eq:bounded identity map}).}\label{algorithm}
$t \gets 0$\; 
\lIf{$A(u) = N$}{ 
    \Return $x^{(0)} \gets u $
    }
\lElseIf{$B(\ell) = N$}{
    \Return $x^{(0)} \gets \ell$}
$\mathcal{P} \gets \text{ the power set of } B(\ell)$\;
\For{$i=0;\ i < |\mathcal{P}|; \ i=i+1$}{
$\hat{u} \gets \ell \Lambda^{P_i} + u (I - \Lambda^{P_i})$, where $P_i \subset B(\ell)$ is the $i$-th element of $\mathcal{P}$\;
$A_t \gets A(\hat{u}) \cap (N \setminus P_i)$\;
\For{$j=0; \ j < n; \ j = j+1$}{
    $t \gets t+1$\;
    \try{}{
    $x^{(t)} \gets \text{ the fixed point of }$ $x = u \Lambda^{A_{t-1}} + \ell \Lambda^{P_i} + \{[u \Lambda^{A_{t-1}} + \ell \Lambda^{P_i} +  x (I -\Lambda^{A_{t-1}} -  \Lambda^{P_i})] W + \e\}(I -\Lambda^{A_{t-1}} -  \Lambda^{P_i})$\;
    }
    \lexcept{singular matrix error}{
   \textbf{break}
    }
    $A_t \gets A(x^{(t)})$\;
    \lIf{$A_t = A_{t-1}$ }{
    \textbf{break}}
    }
\lIf{$f(x^{(t)} W + \e)= x^{(t)}$}{\Return $x^{(t)}$}    
}
\end{algorithm}

\begin{example}
We consider a numerical example to demonstrate Algorithm \ref{algorithm}.
Let the system $(f, W, \e)$ be
\begin{equation}
\begin{split}
&f = \min\{ \max\{ xW + \e, 0\}, u\} \text{ where } u = (5, 10, 10, 8, 10, 10, 6) \\
& W =
\begin{pmatrix}
    0 & 0.4 & 0.15 & 0 & 0.4 & 0.05 & 0 \\
    0.4 & 0 & 0.15 & 0.25 & 0 & 0.2 & 0 \\
    0.3 & 0.1 & 0 & 0.25 & 0.15 & 0.2 & 0\\
    0 & 0 & 0 & 0 & 1 & 0 & 0 \\
    0 & 0 & 0 & 1 & 0 & 0 & 0 \\
    0 & 0 & 0 & 0 & 0 & 0 & 1 \\
    0 & 0 & 0 & 0 & 0 & 1 & 0
\end{pmatrix}\\
& \e = 10^{-5}(2, 1, -1, 3, 2, -1, -2).
\end{split}
\end{equation}
First, compute $B(\ell) = B(0) = \{3, 6, 7 \}$ and $A(u) = \{1, 4, 5, 6, 7 \}$.
Then, the power set $\mathcal{P}(B(\ell))$ has $2^3$ elements.
When $P_i$ in Algorithm \ref{algorithm} equals $\{3, 6, 7\}$, we have $\hat{u} = (5, 10, 0, 8, 10, 0, 0)$ and $A(\hat{u}) \cap (N\setminus P_i) = \{4, 5 \}$. 
The iterations of Algorithm \ref{algorithm} give:
\begin{table}[!htb]
\begin{tabular}{ll}
$x^{(t)}$ & $A_t$        \\ \hline
 $\hat{u} = (5, 10, 0, 8, 10, 0, 0) $ &  \{4, 5 \}\\
 $x^{(1)} = (2.857 \times 10^{-5},  2.143\times 10^{-5}, 0, 8,
 1, 0, 0)$ &  \{4 \}  \\
$x^{(2)} = (2.857 \times 10^{-5},  2.143\times 10^{-5}, 0, 8,
 8.00003, 0, 0)$  &  \{4\}
\end{tabular}
\end{table}

In this case, since $|\mathcal{P}| =2^3$ and $|N\setminus P_i|=4$, the algorithm converges to the equilibrium in at most $2^5$ iterations.
Alternatively, if we use the operator $T$ of (\ref{eq:operator Tx=f(xW + e)}) to iterate
from the upper bound to compute equilibrium $\lim_{m \ra \infty} T^m u$, the iteration time is more than $4.9\times10^5$ given the convergence tolerance $10^{-5}$.
If we iterate from the lower bound $\lim_{m \ra \infty} T^m \ell $, the iteration time is more than $2.3\times 10^5$.
\qed
\end{example}

Alternatively, we can use (nonlinear) programming to solve the problem (\ref{eq:bounded identity map}) as \cite{eisenberg2001systemic}.
Let $g\colon \R^n \ra R$ be a strictly increasing function.
Define the programming problem as 
\begin{equation}\label{linear programming}
    \begin{split}
        &\max_{x \in [\ell, u]} g(x) \\
        &\text{subject to } 0 \leq \max\{x W + \e - x, \ell - x \}.
    \end{split}
\end{equation}
As shown in \cite{eisenberg2001systemic}, the solution to (\ref{linear programming}) is the (almost surely) unique equilibrium.\footnote{
The solution to the following programming problem is also an equilibrium. 
\begin{equation*}
        \min_{x \in [\ell, u]} g(x) \qquad 
        \text{subject to }  \min\{x W + \e - x, u - x \} \leq 0,
\end{equation*} where $g\colon \R^n \ra R$ is strictly increasing.}

\begin{lemma}\label{lemma:linear programming}
Let $f$ follow (\ref{eq:bounded identity map}), $W \geq 0$ be stochastic, and $\e \in \R^n$.
If $g$ is strictly increasing, then any solution to programming problem (\ref{linear programming}) is an equilibrium.
\end{lemma}

\section{Key Player} \label{sec:key player}
In this section, we utilize the unique equilibrium to identify the most influential agent. 
There are many measures or centralities that evaluate the importance scores for agents in a network.\footnote{See \cite{das2018study} for a survey of centralities.}
For instance, in input-output analysis, the output multiplier measures the overall output impact of a sector when it has a dollar-worth increase in final demand, so we can use the output multiplier to identify the most influential production sector, and then policymakers can decide which sector to bail out during recession \citep{miller2009input}. 
In network games, \cite{networkgame:ballester2006s} define the "key player" as the agent that has the highest total impact on the aggregate activity once she is removed from the network.\footnote{
They show that the key player has the highest intercentrality, a measure defined by Katz (Bonacich) centrality.}
We provide a measure for identifying the key player by casting an equilibrium to the steady state of continuous-time dynamics, following and generalizing the control analysis in \cite{sharkey2017control}. 

Observe that the equilibrium of the interaction system (\ref{eq:interation function}) can be interpreted as the steady state of the following continuous-time dynamics:
\begin{equation} \label{eq:continuous dynamics}
\frac{\de x}{\de t} = f(xW + \e) - x   
\end{equation}
where $x \in \R^n$ are economic states and $f(xW + \e)=(f_j (\sum_i x_i w_{ij} + \e_j))_{j=1}^n$ is defined as before.
The \emph{steady state} $x^*$ of (\ref{eq:continuous dynamics}) is such that $f(x^* W + \e) - x^* = 0$.
Suppose that the interaction functions are increasing.
In this dynamics, an agent's equilibrium increases in others' equilibrium $f(xW+ \e)$ and decreases in the amount of itself $x$.
Clearly, if $y$ is the equilibrium of model (\ref{eq:interation function}), then it is the steady state to the continuous-time dynamic system (\ref{eq:continuous dynamics}) that $\de x / \de t = f(y W + \e) - y = 0$.
If Assumption \ref{assumption:contracting} holds and then the equilibrium is unique by Proposition \ref{proposition:contraction_uniqueness}, then the steady state is also unique.

In this section, suppose that $f_i$ is differentiable for all $i$.
For some network including equation (\ref{eq:generalised EN model}), the interaction function is non-differentiable. 
In this case, we can approximate the interaction function by a smooth function without loss of economic meaning. 

Like \cite{sharkey2017control}, we know that the contraction condition in Proposition~\ref{proposition:contraction_uniqueness}, $r(|W|\diag(\b)) < 1$,   also deliveries a stable continuous-time dynamics by Lemma \ref{lemma:continuous dynamics stable}.

Let $\de{x}/ \de t = F(x(t))$ and $x(0) = x^0$ be an autonomous system, where $x(t) \in \R^n$ denotes the state vector, and $F\colon \R^n \ra \R^n$ is a differentiable function of $x(t)$.
Denote $x^*$ as the steady state such that $F(x^*) = 0$.
Recall that the steady states $x^*$ of an autonomous system is
\emph{stable} if for every $\epsilon > 0$ there exists $\d > 0$ such that $\|x(0) - x^*\| < \d$ implies $\|x(0) - x^*\|< \epsilon$ for all $t \geq 0$.
The steady state $x^*$ is \emph{asymptotically stable} if it is stable and there is $\d > 0$ such that $\|x(0) - x^* \| < \d$ implies $x(t) \ra x^*$ as $t \ra \infty$.

\begin{lemma} \label{lemma:continuous dynamics stable}
Suppose that $f_i$ is increasing and continuously differentiable for all $i$, and $r(|W|\diag \b) < 1$.
Then the dynamic system (\ref{eq:continuous dynamics}) is asymptotically stable.
\end{lemma}

Again, by Lemma \ref{lemma:continuous dynamics stable}, if $W$ is non-negative and $\b_i \equiv \psi$, we can reduce the condition for asymptotic stability to $\psi r(W) < 1$.
To have an asymptotically stable system, assume that $r(|W|\diag(\b)) < 1$ in this section. 

Denote the symbol $\circ$ as the Hadamard product such that $s \circ x \coloneqq [s_1 x_1, \dots, s_n x_n]$ for $s, x \in \R^n$.
Define an alternative continuous-time dynamics as 
\begin{equation} \label{eq:continuous dynamics shocks}
    \frac{\de x}{\de t} = F(x, s) \coloneqq f(xW + \e) - s \circ x
\end{equation}
Here, the coefficients $s$ specify the small shocks to the agents.

When $s = \bm 1$, equation (\ref{eq:continuous dynamics shocks}) is equal to the original system.
Let $x^*$ be the steady state for $s = \bm 1$ that $F(x^*, \bm 1)=0$.
Assume that $s = \bm 1$ so that the analysis is around the equilibrium when there is fluctuation to the system.
Following the definition of key player, we remove agent $i$ from the dynamics (\ref{eq:continuous dynamics shocks}) while the others are holding the same. 
The equivalent shock to the removal of $i$ is
\begin{equation*}
    \frac{\p s_i}{\p x_i^*} x_i^*
\end{equation*}
The impact to the other agent $j$'s steady state is then given by:
\begin{equation*}
    C_{ij} = \frac{\de x_j^*}{\de s_i}\frac{\p s_i}{\p x_i^*} x_i^*
\end{equation*}
where $\de x_j^* / \de s_i$ is the extent of change in $j$'s steady state responding to the shocks.
Hence, $C_{ij}$ measures the impact on agent $j$ when the shock is equivalent to the removal of $i$.
With everything else remaining the same, the total impact of the removal of $i$ is equal to $\sigma_i \coloneqq \sum_j C_{ij}$.
We evaluate the total impact around the steady state $(x, s) = (x^*, \bm 1)$ in the following lemma.
Denote $f'_i(x^* W + \e) \coloneqq (f_i'(\sum_{h} x^*_h w_{hi} + \e_i))_i$.

\begin{lemma}\label{lemma:key player measure}
If $f_i$ is differentiable for all $i$ and $r(|W|\diag(\b)) < 1$, then the total impact is $\sigma = \bm 1  \lb I -  \diag{(f'(x^* W + \e)) W^\top}\rb^{-1}  \diag{(x^*)}$.
\end{lemma}

Therefore, the total impact, when agents are removed in the continuous-time dynamics, is determined by the steady state or equilibrium, the network structure $W$, and the derivative of interaction function.
The key player is the agent with the highest $\sigma_i$.
The term $\diag(x^*)$ implies that the larger the agent's equilibrium state is, the higher the measure $\s_i$ is.  
Moreover, the first term, $\bm 1[I - \diag(f'(x^*W + \e)) W^\top ]^{-1}$ implies that the impact measure depends on the interaction behavior $f'(x^*W + \e)$ and the network connections, $W^\top$.
From the Neumann series, we can see that the more interconnected agents tend to have higher impact measure $\s_i$.\footnote{
Neumann series: $[I-A]^{-1} = I + A + A^2 + A^3 + \cdots$ for a matrix $A$.
A highly interconnected agent also tends to have high equilibrium $x^*$.}
Also, $\bm 1[I - \diag(f'(x^*W + \e))W^\top ]^{-1}$ can be view as the authority-based Katz centrality adjusted by the interaction functions $\diag(f'(x^*W + \e))$.
Hence, the measure $\s$ captures either the too-big-to-fail or too-interconnected-to-fail agents.

Note that unlike \cite{NetworkGame:ballester2004s:crime} 
this control analysis does not change the network structure (i.e., $W$ is the same after removing an agent.)
On the other have, \cite{NetworkGame:ballester2004s:crime} assume that the sensitivity matrix is symmetric in the network (\ref{eq:networkgame:best-reply}), while the control analysis in continuous-time system allows arbitrary network structure. 

One issue for the measure $\s$ is that when $\diag{(f'(x^* W + \e) )} W^\top $ has constant column sums, the measure is collapsed to $\s= x^*$, which is just the comparison among equilibria in magnitude.

In many applications, the derivative of the interaction function is constant.
In this case, we can see that the measure $\sigma$ evaluates both effects of the agents' impact on others and the perturbations received from others. We illustrate this property by the following example.

\begin{example}
Given the network $x = \a x W + \bm 1$ for some $\a \in \R$, the equilibrium is $\bm(I-\a W)^{-1}$ and $f'_i \equiv \a$.
The measure for key player $\sigma$ is reduced to $$\sigma = \bm 1  \lp I - \a W^\top \rp^{-1}  \diag{(\bm 1 (I - \a W)^{-1}}).$$
Define the hub-based Katz centrality as $\kappa_h \coloneqq \bm 1 (I-\a W)^{-1} $, and the authority-based Katz centrality as $\k_a \coloneqq \bm 1 (I-\a W^\top)^{-1}$.
Then, the total impact is the element-wise multiplication of hub-based Katz centrality and authority-based Katz centrality.
$$\s = \k_a \circ \k_h.$$
As the explanation in \cite{sharkey2017control}, the hub-based Katz centrality measures the "receiver" property that agents are affected by others, and the authority-based Katz centrality describes the "sender" property that agents influence others.
The agent have high $\s_i$ if either she is influenced significantly by others or she propagates shocks and affects others significantly.

Similarly, the network game (\ref{eq:network game:interbank loan}) of interbank lending market has the equilibrium $x^* = \e \ \diag(\phi) [ I  - W \diag(\phi)]^{-1}$, where $\phi = (\phi_i)$.
The key player measure is
$$\sigma = \lp \, \bm 1 [I -\diag(\phi) W^\top ]^{-1} \, \rp \circ  \lp \, \e \, \diag(\phi) [ I  - W \diag(\phi)]^{-1} \rp$$
It also illustrates the "receiver" and "sender" effect of shock transmission, where the receiver effect is weighted by $\e \diag(\phi)$.
Due to the term $\e \diag(\phi) = \theta I - \diag(c_0)$ with $c_0 = (c_{0,i})$, if a bank $i$ has lower marginal cost $c_{0, i}$ when it has no links to other banks, then it tends to have higher impact measure. 
Overall, the equation implies that the banks with significant impact measure may be either the too-big-to-fail or too-interconnected-to-fail.
\qed
\end{example}

\section{Conclusion}

We show the (almost surely) uniqueness of equilibrium for the generalized and unified network model.
The uniqueness of equilibrium holds if either the interaction functions or sensitivity matrix has a contraction property such that the corresponding spectral radius is less than one.
Alternatively, if the interaction functions are non-expansive and bounded, and the sensitivity matrix is non-convergent (spectral radius of one), then the equilibrium is unique almost surely, given the absolutely continuous shocks. 
Moreover, we demonstrate that if the interaction functions are non-expansive, the boundedness of the interaction functions is essential to determine the existence and uniqueness of equilibrium.
Using this idea, we can compute the equilibrium of the generalized Eisenberg-Noe interbank  network in finite steps by checking which agents are at the bounds.
Lastly, concerning systemic stability, we illustrate a measure for identifying the key players in the unified network by interpreting the equilibrium into the steady state of a continuous-time dynamic system and computing the total impact of removing an agent. 
The measure has the desired properties to evaluate the impact of both receiving and broadcasting perturbations.
Since either the magnitude of economic states in equilibrium or the strength of interconnection affects the measure, it helps identify either too-big-to-fail or too-interconnected-to-fail agents.

%% file: appendix.tex
\appendix
\section{Intuition of Multiple Equilibria} \label{sec:counterexample}


In this appendix, we explain some intuition for Example \ref{example:counter} in Section \ref{sec:generic uniqueness} with $n$ agents that admits multiple equilibria.
We suppose that $\sum_j w_{ij} = 1$ for all $i$ and $\sum_i \e_i = 0$ as Example \ref{example:counter}.
The interaction function $f$ is the bounded identity mapping (\ref{eq:identity map}) broadcasted to the vectors.

The row-sum assumption imply that $W \bm 1^\top = \bm 1^\top$, where $\bm 1$ is a row vector of ones. We also assume that the matrix $W$ is irreducible as Example~\ref{example:counter}.
Suppose that there exists one solution $x$ such that $-M \bm{1} \leq x = x W + \e \leq M \bm{1}$.
Then, we have $ x\bm 1^\top = x \bm 1^\top+ \e \bm 1^\top$ so that it has to be that $\e \bm 1^\top=0$.
By Perron–Frobenius Theorem and irreducibility, the matrix $W$ has a simple left eigenvector $e$ that is strictly positive and satisfies $e W = e$.
Let $y = x + t e$ and $t \in \R$ such that $y \in [-M \bm 1, M\bm 1]$.
Since $y W + \e =teW + xW + \e =te + x= y$, we create another solution $y$.
We can see that the uniqueness may fail with non-zero probability if  Pr$(\sum_i \e_i=0) > 0$.

\begin{remark}\label{remark:E-N model}
Consider the generalised Eisenberg-Noe financial network (\ref{eq:generalised EN model}) in \cite{acemoglu2015systemic} and \cite{acemoglu2015Endogenous} and assume that $\bar p_i \equiv M$ for all $i$. 
The interaction function is 
\begin{equation} \label{eq:N-E model:indicator function}
   f_i(z) = z \mathbbm{1}_{\{0\leq z < M\}}(z) + M \mathbbm{1}_{\{z\geq M\}}(z).
\end{equation}
for all $i$. In words, the interaction function is non-negative and $f_i(z) = 0$ if $z < 0$.
Moreover, the sensitivity matrix satisfies $\sum_{j} w_{ij} = 1$ for all $i$ but not $\sum_h w_{hi}=1$ for all $i$.
Therefore, we cannot apply the theorem of \cite{acemoglu2016network} directly, although their proof can be extended by relaxing this assumption (see Proposition \ref{proposition:generic uniqueness and uniqueness}).
\cite{acemoglu2015systemic} show that the equilibrium is also generically unique for such interaction function and irreducible sensitivity matrix.


However, we see that Example \ref{example:counter} can be easily modified as an example to show that it may admit multiple equilibria with non-zero probability under the interaction function (\ref{eq:N-E model:indicator function}).
The shock in \cite{acemoglu2015systemic} is equal to the sum of holding cash, project return, and the liquidation of asset minus senior liability (see Section \ref{section:Financial}).
In their setting, the cash, senior liability, and liquidation of asset are all constant, while the realizations of project returns are i.i.d, and only take two values. Hence, the shocks are also i.i.d. and $\e_i \in \{e_1, e_2\}$ with $e_1 > 0$ and $e_1 < 0$.\footnote{The holding cash and liquidation are both zero in section three of \cite{acemoglu2015systemic}.}
Therefore, it is possible that uniqueness fails with non-zero probability when we assign strictly positive probabilities to both $e_1$ and $e_2$.
Hence, the generic uniqueness result in \cite{acemoglu2015systemic} may be problematic when the shock variable is not absolutely continuous.


\qed
\end{remark}

\begin{remark}
\cite{hurd2016contagion} shows that the multiple solutions of generalized Eisenberg-Noe model (\ref{eq:generalised EN model}) are eigenvectors of $W$. 
Clearly, Example \ref{example:counter} and the above discussion show that the claim is incorrect, and it should be that the differences in solutions are eigenvectors.
\qed
\end{remark}

\section{Proofs in Section \ref{sec:generic uniqueness}} 
\label{appendix}

The proofs of Proposition \ref{proposition:contraction_uniqueness} and proposition \ref{proposition:generic uniqueness and uniqueness} follows the subsequent lemmas.
To begin with, recall that we have the map $T\colon \R^n \ra \R^n$:
\begin{equation} \label{eq:operator T}
    T x \coloneqq f\left(x W + \e \right)
\end{equation}
for $x \in \R^n$, where $f(xW+\e) = (f_i(\sum_h x_h w_{hi} + \e_i))_{i\in N}$.
Clearly, the vector $x$ is a fixed point of $T$ if and only if it is an equilibrium.

\begin{proof}[Proof of Lemma \ref{lemma:existence}]
Suppose that $f_i$ is increasing and bounded for all $i \in N$.
Then, there is $E_i >0$ such that $|f_i(t)| < E_i$ for all $i$.
Let $E = (E_i)$. 
Define $T$ by (\ref{eq:operator T}).
Hence, it must that $Tx \in [-E, E]$ for all $x \in [-E, E]$.
Note that if $x^* > E$ or $x^* < -E$ is a fixed point of $T$, then we have $Tx^* \leq E < x^*$ or $x^* < -E \leq Tx^*$, which contradicts the fact that $x^*$ is a fixed point.
Hence, we can restrict the domain of $T$ on $[-E, E]$ without loss of generality, so that $T$ is an increasing self-map on $[-E, E]$.
Tarski's Fixed Point Theorem shows that there exists an equilibrium in $[-E, E]$,  and the set of equilibria forms a complete lattice, whence both the highest and the lowest equilibria exist.
\end{proof}

The following lemma shows that $T$ is a contraction mapping if $\| \lp |W |\diag{(\b)} \rp^k\| < 1$.

\begin{lemma}\label{lemma:T^k contraction}
$\| T^k x - T^k\hat x\| \leq \| \,\lp \, |W | \diag{(\b)} \rp^k\|  \|x - \hat x\|$ for all $x, \hat x$ in $\R^n$ and $k \in \N$.
\end{lemma}
\begin{proof}[Proof of Lemma \ref{lemma:T^k contraction}]
Let $x, \hat x \in \R^n$ and define $T$ by (\ref{eq:operator T}).
We first show that $|T^k x - {T}^k \hat x| \leq |x-\hat x| \, (|W| \diag{(\b)})^k $ for all $k \in \N$ by induction.
Since $Tx = f(xW + \e)$ and $|f(xW + \e) - f(x|$, we have

\begin{equation*}
\begin{split}
\left|Tx - T\hat x\right| &\leq  \left|(xW + \e) - \left(\hat x W  + \e\right) \right|\diag{(\b)}\\ 
&= \left|(x - \hat x)  W \right| \diag{(\b)} \leq |x - \hat x| |W| \diag{(\b)}.   
\end{split}
\end{equation*}
The claim holds for $k = 1$.
Suppose that the claim holds for some $m \in \N$, i.e. $|T^mx - T^m \hat x| \leq | x - \hat x|  (|W|\diag{(\b)})^m $.
By iteration, we have
\begin{equation*}
\begin{split}
    \left|T^{m+1} x- T^{m+1} \hat x \right|&= \left |f\left(T^m x W + \e\right) - f\left(T^m \hat x W+ \e \right)\right|\\
     &\leq \left|T^m x  W-  T^m \hat x W\right| \diag{(\b)}\\
    & \leq \left| T^m x - T^m \hat x \right| | W | \diag{(\b)}\\
    &\leq   |x- \hat x|\, (|W| \diag{(\b)} )^{m+1}.
\end{split}
\end{equation*}
Thus, the induction implies that $|T^k x - T^k \hat x| \leq | x - \hat x| (|W |\diag{(\b)} )^k $ for all $k \in \N$. 
The definition of $p$-norm
gives $\|T^k x - T^k \hat x\| \leq \|| x - \hat x| (|W| \diag{(\b)} )^k \| \leq  \| ( |W | \diag{(\b)})^k\| \, \| x - \hat x\|$ for all $k \in \N$.\footnote{For the max norm, we can take the maximum on both sides of inequality.
For a \emph{Riesz norm},  $|x| \leq |y|$ implies $\|x\| \leq \|y\|$.}
\end{proof}
We are ready to prove the first statement of Proposition \ref{proposition:contraction_uniqueness} and its following corollaries.

\begin{proof}[Proof of Proposition \ref{proposition:contraction_uniqueness}]
Let $f$ and $W$ satisfy Assumption \ref{assumption:contracting} so that $r(|W|\,\diag{(\b)}) < 1$, where $\b = (\b_j)$.
Then, we have $$\lim_{k \ra \infty} \|(|W|\,\diag{(\b)})^k \| = 0.$$ Thus, $\| (\, |W|\,\diag{(\b)})^k\| < 1$ for some $k \in \N$.
Lemma \ref{lemma:T^k contraction} and the Banach Contraction Theorem show that there is a unique equilibrium.
\end{proof}

\begin{proof}[Proof of Corollary \ref{corollary:globally stable}]
Let $f, W$ and $\e$ satisfy the conditions of Proposition \ref{proposition:contraction_uniqueness}.
Let $x^*$ be the equilibrium and $T\colon \R^n \ra \R^n$ be defined as $Tx = f(xW + \e)$.
Denote $A = |W| \diag{(\b)}$.
The first statement follows from Proposition \ref{proposition:contraction_uniqueness} and Banach Fixed Point Theorem.
From Lemma \ref{lemma:T^k contraction}, we have $\| T^{k+1} x  - T^k x \| = \|T^k (T x) - T^k x \| \leq \|A^k \| \|T x - x\| $ for all $k \in \N$.
\end{proof}

\begin{proof}[Proof of Corollary \ref{corollary: r(W diag(b))<1}]
Let $f$ and $W$ satisfy the conditions in Corollary \ref{corollary: r(W diag(b))<1} so that $r(W\,\diag{(\b)}) < 1$, where $\b = (\b_j)$.
The proof is similar to Lemma \ref{lemma:T^k contraction} and Proposition \ref{proposition:contraction_uniqueness}.
In detail, we can show that $T^k x - T^k \hat x = (x-\hat x) \, (W \diag{(\b)})^k $ for all $k \in \N$ by the same induction in Lemma \ref{lemma:T^k contraction}.
Thus, we have $\|T^k x - T^k \hat x \| \leq \|x-\hat x\|  \, \| (W \diag{(\b)})^k \|$.
If $r(W\diag(\b)) < 1$, then $\| (W \diag{(\b)})^k \| < 1$ for some $k$.
The result follows the Banach Contraction Theorem.
\end{proof}

\begin{proof}[Proof of Lemma \ref{corollary:convergence condition}]
Let $f$ and $W$ be such that the conditions of Lemma \ref{corollary:convergence condition} hold.
Let $\b=(\b_i)$ be Lipschitz constants.
If the first condition of Lemma \ref{condition:convergence} holds, it implies that $\|W^m\|_\infty < 1$ for some $m \in \N$ by Theorem 2.5 of \cite{azimzadeh2019contraction}.
Similarly, the second condition of Lemma \ref{condition:convergence} implies that $\|W^m\|_1 = \|(W^\top)^m\|_\infty < 1$ for some $m\in \N$.
Since $\diag(\b )= I$ for non-expansiveness of $f$, we have $r(|W|\diag(\b)) = r(W) < 1$. 
\end{proof}

\begin{proof}[Proof of Lemma \ref{corollary:acyclic}]
Let $f$ and $W$ be such that the conditions of Lemma \ref{corollary:acyclic} hold.
Let $\b=(\b_i)$ be Lipschitz constants.
An acyclic graph contains some sink node $i$ such that $\sum_{j} w_{ij}=0$. 
For each node $s$ in an acyclic graph, there is an acyclic path to some sink node $t$.
Suppose that node $s$ can reach $t$ in $K_t$ steps such that $w_{st}^{K_t} > 0$. 
Let $K^* = \max_t K_t$ be the length from node $s$ to the farthest connected sink node $t^*$.
(Since if there are multiple farthest sink nodes, they have the same length. 
Without loss of generality, assume that there is only one farthest sink node from $s$.) 
Then, we must have $ \sum_u w_{s u}^{K^*+1} = 0$.\footnote{If not zero, it contradicts with the definition of $K^*$.}
Since $s$ is arbitrary, we have $W^T = 0$ for some large enough $T \in \N$. It implies that $r(W) = 0.$
Let $A = |W| \diag(\b)$.
The sink nodes for $\graph{W}$ are also the sink nodes for $\graph{A}$.
Hence, the same argument shows that $r(A)=0$.
\end{proof}


\begin{proof}[Proof of Lemma \ref{lemma:substochasitc_irreducible}]
Let $f$ and $W$ be such that the conditions of Lemma \ref{lemma:substochasitc_irreducible} hold.
Then, the conditions of Lemma \ref{condition:convergence} are satisfied.
The result follows from Lemma \ref{corollary:convergence condition}.
\end{proof}


Next, we attempt to prove the almost surely uniqueness for the case with non-negative $f_i$ and matrix with spectral radius one.
Firstly, we consider that $W$ is irreducible or the $\graph W$ is strongly connected.
Note that \cite{acemoglu2016network} use the continuity of $f$ to show the existence of equilibrium. 
However, they argue that $f$ has at most countably many discontinuous points when they show the uniqueness of equilibrium.
Since $f$ is non-expansive, it is Lipschitz continuous so that their argument is ambiguous.
We provide an alternative proof following \cite{acemoglu2016network}.

\begin{lemma}[\cite{acemoglu2016network}] \label{lemma:unit slope interval}
If $f$ is increasing and $|f(a) - f(b)| = |a - b|$ for $a, b \in \R$ and $b > a$, then $f$ is linear in the interval $[a, b]$ with a unit slope, i.e., $f(z) = z + f(a) - a$ for all $z \in [a, b]$.\footnote{See lemma B.2 of \cite{acemoglu2016network}.}
\end{lemma}

\begin{lemma} \label{lemma:finite linear map}
If $f$ is increasing and bounded, there are at most countably many intervals that $f$ is linear with unit slopes in these intervals.
\end{lemma}
\begin{proof}
Let $f$ be an increasing and bounded mapping.
Let $I$ be the set of intervals such that $f$ is linear and with unit slopes in these intervals.
Let $[a_i, b_i] \in I$. 
Since $b_i > a_i$ and $f$ is an increasing and linear mapping on $[a_i, b_i]$ with a unit slope, we have $f(a_i) <  f(b_i)$.
Then, there is a rational $c_i \in \Q $ such that $f(a_i) < c_i < f(b_i)$.
The map $[a_i, b_i] \mapsto c_i$ is injective, so the set $I$ is countable.
\end{proof}                                             

\begin{lemma}  \label{lemma:irreducible, generically unique}
Suppose that $f_i$ is increasing, non-expansive and bounded for all $i$, the shock is absolutely continuous for all $i$, and $W$ is irreducible with $r(W) = 1$. Then, the equilibrium is almost surely unique. 
\end{lemma}
\begin{proof}[Proof of Lemma  \ref{lemma:irreducible, generically unique}]
Let $f=(f_i)$, $W$ and $\e$ be such that the conditions of Lemma \ref{lemma:irreducible, generically unique} hold.
By Lemma \ref{lemma:existence} or Tarski's Fixed Point Theorem, the equilibrium exists and the set of equilibria is complete.
Without loss of generality, assume that there are two different equilibria $x$ and $y$ with $x > y$.
Since $f$ is non-expansive, we have
\begin{equation} \label{equation:inequality of irreducible W}
\begin{split}
 e \coloneqq  x - y &= |f(x W + \e) - f(yW + \e)| \\
 &\leq |xW + \e - (yW+\e)| = | (x-y)W| = (x-y)W = eW.   
\end{split}
\end{equation}
Let $q \in \R^n$ be the right eigenvector corresponding to $r(W)$ so that $W q^\top  = r(W) q^\top$.
By the Perron-Frobenius Theorem and irreducibility, $q$ is strictly positive, $q \gg 0$. 
Suppose that the inequality of equation (\ref{equation:inequality of irreducible W}) does not bind. 
It implies that
$$e q^\top< e W q^\top = e q^\top,$$
which is a contradiction.
Hence, it must be that the equality of (\ref{equation:inequality of irreducible W}) binds so that $e = |f(xW + \e) - f(yW + \e)| = |(x-y)W| = eW$.
We see that the vector $e$ is the left eigenvector of $W$, so Perron-Frobenius Theorem shows that $e \gg 0$.
Now, define $b \coloneqq xW + \e$ and $a \coloneqq yW + \e$.
Then, we have $|f_i(b_i) - f_i(a_i)| = |b_i - a_i|$ for all $i \in N$.
Since $b - a = eW = e \gg 0$, we have $b \gg a$.
Therefore, the interval $[a_i, b_i]$ is well-defined for all $i$.
Since $f_i$ is increasing for all $i$, it follows from Lemma \ref{lemma:unit slope interval} that $f_i$ is a linear mapping with unit slopes in the intervals $[a_i, b_i]$ for all $i$.
That is, we get $f(z) = z + f(a) - a$ for all $z \in [a, b]$.
In particular, since $x$ is the fixed point, we have $x = f(b) = b + f(a) - a = x W + \e +f(a) - a$. 
Multiplying $q^\top$ on both sides, it implies that $xq^\top = x W q^\top + \e q^\top + (f(a) - a)q^\top = x q^\top + \e q^\top + (f(a) - a)q^\top$.
Since $q$ is the right eigenvector, if there are multiple fixed points, it has to be that $\e q^\top= (a - f(a))q^\top$.
Denote the sets $E$ and $A$ as
\begin{align*}
E \coloneqq \{\e \in \R^n: & \text{There are multiple equilibria.} \} \\ 
A\coloneqq \{a \in \R^n:& \text{$f_i$ is a linear map with unit slope on $[a_i, b_i]$ } \\ 
& \qquad \qquad \qquad \text{with $b_i > a_i$ for $i=1, \dots, n$.}  \}.
\end{align*}
By the above discussion, we have $\e \in E$ only if $ \e q^\top= (a - f(a))q^\top$, so $E \subset E_1 \coloneqq \{\e \in \R^n: \e q^\top = (a - f(a))q^\top, a \in A \}$.
Therefore, $\e_1, \dots, \e_n$ are linearly dependent when there exists multiplicity.
Since Lemma \ref{lemma:finite linear map} implies that there are countably many intervals $[a_i, b_i]$ for $f_i$, there are countably many vectors $a$, whence $A$ is at most countable.
Therefore, the dimension of $E_1$ is strictly less than $n$, and the Lebesgue measures of $E_1$ and $E$ are zero.
Since the measure of shocks is absolutely continuous, the probability of realizations in $E$ is zero, $\mathbbm{P}_\e(\e \in E)=0$.
\end{proof}

Define in-subgraph and out-subgraph for the proof of Proposition \ref{proposition:generic uniqueness and uniqueness}.
Let $A = (a_{ij}) \in \R^{n\times n}$ and $\graph A = (V, E)$. 
Let $\graph S = (V_S, E_S)$ be a \emph{subgraph} of $\graph A$ such that $V_S \subset V$ and $E_S \subset E$.
Subgraph $S$ is an \emph{in-subgraph} if there is no path from $i$ to $j$ for some $i \in V_S$ and $j \in V\setminus V_S$.
Subgraph $S$ is an \emph{out-subgraph} if there is a path from $i$ to $j$ for some $i \in V_S$ and $j \in V\setminus V_S$.

\begin{proof}[Proof of Proposition \ref{proposition:generic uniqueness and uniqueness}]
Let $f=(f_i)$ and $W$ be such that the conditions of Assumption \ref{assumption:non-expansive} hold, and $\e$ be absolutely continuous.
Suppose that $r(W) = 1$.
By Lemma \ref{lemma:existence}, the equilibrium exists.
Define the equivalence relation $\sim$ as $i \sim j$ if and only if $i$ and $j$ are accessible from each other for all $i, j \in N$.
Denote $[i]\coloneqq \{s \in N: s \sim i \}$ as the equivalence class. 
We see that $[i] \neq [j]$ if either $i$ is not accessible from $j$, or $j$ is not accessible from $i$, so the equivalence classes are mutually disjoint.
Moreover, the set $\{[i]: i \in N \}$ forms a partition of $N$.
By definition, each equivalence class $[i]$ is either an in-subgraph or out-subgraph.
Since the number of verteces is finite, there are finite in-subgraphs, denoted by $S_1, \dots, S_m$.
Let $S_0$ be the union of all out-subgraphs.
By definition of in-subgraphs and out-subgraphs, we can permute the matrix $W$ and write it as \footnote{See also \cite[chapter 8]{berman1994nonnegative}.}
\[W= 
\begin{pmatrix}
W_0 & P_1 & P_2& \cdots & P_m \\
0  & W_1  & 0 & \cdots & 0 \\
0  & 0 & W_2 & \cdots & 0 \\
 \vdots & \vdots & \vdots & & \vdots \\
0 & 0& 0& \cdots & W_m\\
\end{pmatrix}\]
where $W_t \in \R^{|S_t|\times |S_t|}$ is the weighting adjacency matrix of subgraph $S_t$, and $P_t \in \E^{|S_0|\times|S_t|}$ describes the path from $S_0$ to $S_t$.\footnote{$|S_t|$ denotes the number of nodes in subgraph $S_t$.}
Note that $W_1, \dots, W_m$ are irreducible, since in-subgraphs are strongly connected.
Let $x$ be an equilibrium of states.
Decomposing the state vector and shock vector into block forms corresponding to $S_0, S_1, \dots, S_m$, we have $x = (x_0, x_1, \dots, x_m)$ and $\e=(\e_0, \e_1, \dots, \e_m)$, where $x_t, \e_t \in \R^{|S_t|}$ for $t = 0, 1,\dots, m$.
The equilibrium system can be written as
\begin{equation*}
\begin{split}
x_0 &= f(x_0 W_0 + \e_0) \\
x_1 &= f(x_1 W_1 + x_0 P_1 + \e_1)\\
x_2 &= f(x_2 W_2 + x_0 P_2 + \e_2)\\
& \, \ \vdots\\
x_m &= f(x_m W_m + x_0 P_m + \e_m)
\end{split}
\end{equation*}
where $f(x_t W_t + x_0 P_t + \e_t) \coloneqq (f_i(\sum_{h \in S_t} x_h w_{hi} + \sum_{h \in S_0} x_h w_{hi} + \e_i))_{i \in S_t}$.
Since $r(W) = 1$, we have $r(W_0) \leq 1$.
We first assume that $r(W_0)<1$.
Then, $x_0 = f(x_0 W_0 + \e_0)$ has a unique solution following Proposition \ref{proposition:contraction_uniqueness}.
Fix $t = 1, \dots, m$.
Given the unique $x_0$, Lemma~\ref{lemma:irreducible, generically unique} shows that the equilibrium $x_t$ is almost surely unique.
It is not unique only if $ (x_0 P_t + \e_t)q_t^\top = (a_t - f(a_t))q_t^\top$ from Lemma~\ref{lemma:irreducible, generically unique}, where $q_t$ is the right eigenvector for $W_t$ and $f_i$ is a unit-sloping mapping in intervals $[a_{t,i}, b_{t,i}]$ for all $i\in S_t$.
Denote the set 
\begin{equation*}
E_t\coloneqq \{\e \in \R^n: \text{the system } x_t = f(x_t W_t + x_0 P_t + \e_t) \text{ has multiple solutions} \}.
\end{equation*}
We see that $E_t$ has Lebesgue measure zero by the argument in Lemma~\ref{lemma:irreducible, generically unique}.
It implies that $\lambda(E_1\cup \cdots \cup E_m)\leq \lambda(E_1) + \dots +\lambda(E_m) =0$, where $\lambda$ denotes the Lebesgue measure.
Let $E = \bigcup_t E_t$.
Since the random variable of $\e$ is absolutely continuous, we have $\mathbbm{P}_\e(\e \in E) = \int_E g(x) \de \lambda = 0$, where $g$ is the density function of shock.
Overall, each in-subgraph has almost surely unique equilibrium and the equilibrium $x$ is also almost surely unique.

Next, suppose that $r(W_0) = 1$, and $W_0$ is irreducible.
Denote the equilibrium of $x_0 = f( x_0 W_0  + \e_0)$ as $x_0(\e_0)$.
In this case, the solution $x_0(\e_0)$ is unique almost surely by Lemma \ref{lemma:irreducible, generically unique}.
If $x_0$ is unique, the equilibrium $x$ is almost surely unique following the above argument.
If $x_0$ is not unique, then it must be $ \e_0 q_0^\top = (a_0 - f(a_0)) q_0^\top$, where $q_0$ is the right eigenvector of $W_0$ and $a_0$ is some vector in $R^{|S_0|}$ by Lemma \ref{lemma:irreducible, generically unique}.
When $x_0$ is not unique, the solution $x_t$, for $t=1,\dots, m$, is not unique only if $ (x_0(\e_0) P_t + \e_t)q_t^\top = (a_t - f(a_t))q_t^\top$.
Note that $\{\e \in \R^n \colon  \e_0 q_0^\top = (a_0 - f(a_0)) q_0^\top \text{ and } (x_0(\e_0) P_t + \e_t)q_t^\top = (a_t - f(a_t))q_t^\top \}$ is a subset of $\{\e \in \R^n\colon \e_0 q_0^\top = (a_0 - f(a_0)) q_0^\top\}$, which is measure zero.
Therefore, $x_t$ is also almost surely unique for all $t=1, \dots, m$.

Suppose that $r(W)=1$ and $W_0$ is not irreducible. 
Considering an equivalence class as an entity, since $S_0$ is the collection of out-subgraphs of equivalence class $[i]$, these equivalence classes forms an acyclic graph, otherwise they will deviate their definition.
Assume that there are $r$ equivalence classes $[i]$ in $S_0$.
Since $W_0$ forms an acyclic network, $W_0$ can be written as a upper triangular matrix as
\[W_0 = 
\begin{pmatrix}
B_{11} & B_{12} & B_{13} & \cdots & B_{1r} \\
0 & B_{22} &  B_{23}& \cdots & B_{2r} \\
0 &  0 & B_{33} & \cdots & B_{3r}\\
\vdots & \vdots & & & \vdots \\
0 & 0& 0& \cdots & B_{rr}
\end{pmatrix}\]
where $B_{tt}$ is a square matrix with respect to the corresponding equivalence class for $t = 1, \dots r$.
Then, we can decompose $x_0 = f(x_0 W_0 + \e_0)$ into $r$ systems and pin down the equilibrium in orders.
Since each equivalence class is strongly connected, each $\text{subgraph }B_{tt}$ is irreducible.
Using the similar argument above, Lemma \ref{lemma:irreducible, generically unique} and Proposition \ref{proposition:contraction_uniqueness}, we see that the equilibrium for the system with respect to $\graph B_{11}$ is almost surely unique solution if $r(B_{11}) = 1$ and unique if $r(B_{11}) < 1$.
For $t=2,\dots, r$, if $r(B_{tt})$ is less than one, the system with respect to $B_{tt}$ has a unique equilibrium.
If the $r(B_{tt})$ is one, its equilibrium is almost surely unique.
Overall, $x_0 = f(x_0 W_0 + \e_0)$ has almost surely unique equilibrium following the above argument.
The similar argument for the irreducible case concludes the result for the whole system $x = f(xW + \e)$.
\end{proof}

 \begin{proof}[Proof of Corollary \ref{corollary:iteration from above}]
Suppose that $(f, W, \e)$ follows the assumptions of Proposition \ref{proposition:generic uniqueness and uniqueness}.
Define $T: [\ell, u] \ra [\ell, u]$ as $Tx \coloneqq f(xW + \e)$ following the statement.
Hence, since $u$ is the upper bound, we get $T u \leq u$.
Since $T$ is monotone given the monotonicity of $f$, we get $T^2 u \leq Tu$, whence the iteration implies that $T^n u \leq T^{n-1} u \leq \cdots \leq Tu \leq u$.
Let $x_n \coloneqq T^n u$ for all $n \in \N$.
Then, $(x_n)$ is a non-increasing sequence and bounded below by $\ell$, so it has a limit, denoted by $x^* \coloneqq \lim_{n \ra \infty}T^n u$.
Now, consider $ T x_n = T^{n+1} u$.
Letting $n \ra \infty$ on both sides of the equation, since $T$ is continuous from above given the continuity of $f$, we have $T x^* = T(\lim_{n \ra \infty} x_n) = \lim_{n \ra \infty} T x_n = \lim_{n \ra \infty}  T^{n+1} u = x^*$.
Therefore, $x^*$ is the fixed point of $T$. 
Since Proposition \ref{proposition:generic uniqueness and uniqueness} implies that $T$ has a unique fixed point almost surely, $x^*$ is the almost surely unique equilibrium. 
\end{proof}

\begin{proof}[Proof of Corollary \ref{corollary:multiplicity subgraph}]
Let $f$ and $W$ be such that Assumption \ref{assumption:non-expansive} holds.
Using Lemma \ref{corollary:acyclic} and Proposition \ref{proposition:contraction_uniqueness}, if the network graph is acyclic, then the equilibrium is unique.
Hence, if there exists multiplicity, there must be a strongly connected subgraph.
We then decompose the graph into strongly connected subgraphs as the proof of Proposition \ref{proposition:generic uniqueness and uniqueness}.
The proof of Proposition \ref{proposition:generic uniqueness and uniqueness} implies that, given the realized shocks $\e$, the multiplicity occurs in and originates from some strongly connected in-subgraphs or out-subgraphs $S$ (the shocks corresponding to $S$ are linearly dependent).
It implies that all agents in $S$ admit multiple equilibria.\footnote{
Since $S$ admit multiple equilibria (shocks are linearly dependent), given an equilibrium,  we can create another equilibrium.  
In particular, let $W_s$ be the submatrix of $W$ corresponding to subgraph $S$, and $x_s$ be the equilibrium for agents in $S$.
From equation (\ref{equation:inequality of irreducible W}), $x_s + t e_s W_s$ is another equilibrium, where $e_s$ is the eigenvector of $W_s$, and $t\in \R$ is a parameter such that $x_s + t e_s W_s$ is in the range of the interaction functions.}
More, the agents with multiple equilibria must be accessible from at least one subgraph which has multiple equilibria.
\end{proof}

\begin{proof}[Proof of Corollary \ref{corollary:EN uniqueness}]
Let $f$, $W$ and $\e$ follow the conditions of Corollary \ref{corollary:EN uniqueness}.
Let $\bar{p} = (\bar{p}_i)$.
By Lemma \ref{lemma:existence}, we know that the equilibrium exists.
From the proof of Proposition \ref{proposition:generic uniqueness and uniqueness}, if there are multiple equilibria, there must be some irreducible subgraph of $W$ that admit multiple equilibria.
Without loss of generality, assume that $W$ is irreducible.
On the contrary, suppose that $x$ and $y$ are two fixed points with $x > y$.
By equation (\ref{equation:inequality of irreducible W}) and Lemma \ref{lemma:irreducible, generically unique}, we know that it must follow $|x - y| = |f(xW + \e) - f(yW + \e)|$.
Hence, we have $|x - y| = |(xW + \e)^+ \land \bar{p} - (yW + \e)^+ \land \bar{p} |$.
Since $|(\a)^+ \land \g  - (\b)^+ \land \g| = |\a - \b|$ for $\g > 0$ and $\a, \b \in \R$ if and only if $\a, \b  \in [0, \g]$, we get $0 \leq xW + \e, yW + \e \leq \bar{p}$.
Therefore, we have $x = f(xW + \e) = xW + \e$ and $y = f(yW + \e) = yW + \e$.

By the Perron-Frobenius theorem, the right eigenvector $q$ of $W$ is strictly positive.
Since the row sum is one, $q$ is a vector of ones or constants.
Multiplying both side of $x = xW  +\e$ by $q^\top$, we have $x q^\top = (xW + \e)q^\top = x q^\top + \e q^\top$.
Thus, we get $\e q^\top=0$, whence $\e \bm{1}^\top=0$.
Since $\e \geq 0$ and there is $\e_i > 0$ for some $i$, we obtain a contradiction.
\end{proof}

\section{Proofs in Section \ref{sec:tightness and boundedness}} 

 \begin{proof}[Proof of Lemma \ref{lemma:comparative:contracting}]
Suppose that $(f, W, \e)$ and $(f', W', \e')$ are two networks such that Assumption \ref{assumption:contracting} holds.
Assume that $W \leq W'$, $\e \leq \e'$, $f_i$ and $f_i'$ are increasing functions for all $i \in N$, and $f_i(t) \leq f'_i(t)$ for all $t \in \R$ and all $i \in N$.
It then follows from Proposition \ref{proposition:contraction_uniqueness} that $(f, W, \e)$ and $(f', W', \e')$ have unique equilibrium $\hat x$ and $\hat x'$, respectively.
Define the maps $T\colon \R^n \ra \R^n$ and $\hat T\colon \R^n \ra \R^n$ as $T x \coloneqq f(xW + \e)$ and $\hat T x \coloneqq f'(x W' + \e')$ for all $x \in \R^n$.
We first show that $T^k x \leq \hat T^k x$ for all $x \in \R^n$ and $k \in \N$.
Since $f_i(t) \leq f'_i(t)$ for all $t \in \R$ and all $i$, we see that $f(x) \leq f'(x)$ for all $x \in \R^n$.
Then, since $f_i$ and $f_i'$ are increasing functions for all $i$, and $W \leq W'$, and $\e \leq \e'$, we obtain $Tx =f(xW + \e) \leq f(xW' + \e') \leq f'(xW' + \e') = \hat T x$ for all $x \in \R^n$.
Suppose the induction hypothesis that $T^k x \leq \hat T^k x$ for some $k \in \N$ and all $x$.
Hence, since $Tx \leq \hat T x$ for all $x \in \R^n$, we have $T (T^k x) \leq  \hat T (T^k x) \leq \hat T (\hat T^k x)$.
Therefore, the Mathematical Induction implies that $T^k x \leq \hat T^k x$ for all $x \in \R^n$ and all $k \in \N$.
Since Corollary \ref{corollary:globally stable} shows that $\hat x = \lim_{k \ra \infty} T^k x$ and $\hat x' = \lim_{k \ra \infty} \hat T^k x$ for any $x \in \R^n$,
we obtain $\hat{x} = \lim_{k \ra \infty} T^k x \leq \lim_{k \ra \infty} \hat{T}^k x = \hat{x}'$.
\end{proof}

 \begin{proof}[Proof of Lemma \ref{lemma:comparative:non-contracting}]
Let $(f, W, \e)$ and $(f', W', \e')$ be two networks such that the assumptions of the statement hold.
Suppose that $f_i(t) \leq u_i$ and $f'_i(t) \leq u'_i$ for all $t \in \R$ such that $u_i \leq u_i'$ for all $i$.
Let $u \coloneqq (u_i)$ and $u' \coloneqq (u_i')$.
Define the maps $T\colon \R^n \ra \R^n$ and $\hat T\colon \R^n \ra \R^n$ as $T x \coloneqq f(xW + \e)$ and $\hat T x \coloneqq f'(x W' + \e')$ for all $x \in \R^n$.
Following Proposition \ref{proposition:generic uniqueness and uniqueness}, we know that the equilibrium for both $(f, W, \e)$ and $(f', W', \e')$ are unique almost surely, denoted by $\hat x$ and $\hat x'$, respectively.
Moreover, it follows from Corollary \ref{corollary:iteration from above} that $\hat x = \lim_{k \ra \infty} T^k u$ and $\hat x' = \lim_{k \ra \infty} \hat T^k u'$.
Now, we show that $T^k u \leq \hat T^k u'$ for all $k \in \N$.
Since $f_i(t) \leq f_i'(t)$ for all $t$ and all $i$, we see that $Tx = f(xW + \e) \leq f'(xW + \e) \leq f'(xW' + \e') = \hat T x$ for all $x$.
Hence, since $u \leq u'$, we have $ T u \leq \hat T u \leq \hat T u'$.
Suppose that $T^k u \leq \hat T^k u'$ for some $k \in \N$.
Then, since $T x \leq \hat T x$ for all $x$, and $\hat T$ is an increasing map, we get $T T^k u \leq \hat T T^k u \leq \hat T \hat T^k u'$.
Therefore, the Mathematical Induction implies that $T^k u \leq \hat T^k u'$ for all $k \in \N$.
Consequently, we have $\hat x = \lim_{k \ra \infty} T^k u \leq \lim_{k \ra \infty} \hat T^k u' = \hat x'$.
\end{proof}

\section{Proofs in Section \ref{sec:tightness and boundedness}}

Define the set of $\e$ that the solution exists as 
$$E_0 = \{\e \in \R^n: \text{Linear system (\ref{eq:linear system}) has a solution.} \}$$
Define the set of $\e$ that allows multiple solutions as 
$$E_1 \coloneqq \{\e \in \R^n: \text{Linear system (\ref{eq:linear system}) has multiple solutions.} \}$$

\begin{lemma} \label{lemma:E measure zero}
If $W$ is non-negative, and $r(W)=1$, then $\lambda(E_0) = 0$ and $E_0 = E_1$.
\end{lemma}
\begin{proof}[Proof of Lemma \ref{lemma:E measure zero}]
Let $W$ be a non-negative, irreducible matrix $W$ with $r(W)=1$.
Using Perron-Frobenius Theorem, $r(W)$ is an eigenvalue of $W$, and its corresponding left eigenvector and right eigenvector are non-negative, denoted by $v$ and $e$, respectively.
First, note that $x = xW + \e$ has a solution if and only if $\e e^\top = 0$.
To see this, let $A\coloneqq I - W$ and rewrite the linear system (\ref{eq:linear system}) as $x A = \e$.
Hence, if $x$ is a solution, then $x A e^\top = x e\top - x W e^\top = x e^\top - x e^\top = 0 =  \e e^\top$.
Conversely, if $\e e^\top = 0$, then since $A e^\top = e^\top - e^\top = 0$, we have $rank(A) = rank([A| \e])$, where $[A|\e]$ is the augmented matrix for $xA = \e$.
Since the solution exists if and only if $rank(A) = rank([A| \e])$, there must be a solution.
Therefore, $E_0 = \{\e: \e e^\top=0 \}$ is measure zero. 
For the second statement, it is clear that $E_1 \subset E_0$.
Suppose that $\e \in E_0$. 
Thus, there is $x$ such that $x = xW + \e$.
Letting $t \in \R$, since $v$ is the left eigenvector, we have $x + t v = xW +  \e + tv = (x+tv)W + \e$, whence $x+ tv$ is a solution.
Then, $\e \in  E_1$ and $E_0 \subset E_1$.
\end{proof}

\begin{proof}[Proof of Lemma ~\ref{lemma:linear:existence}]
Let $W$ and $\e$ follow the conditions of Lemma ~\ref{lemma:linear:existence}.
Lemma \ref{lemma:E measure zero} and absolute continuity of shock imply that $\mathbbm{P}_\e(\e \in E_0 )=0$.
\end{proof}

\begin{proof}[Proof of Lemma \ref{lemma:unique solution at boundedness}]
Let $\ell, u \in \R^n$ be such that  $u \gg \ell$, $f=(f_i)$ be defined as (\ref{eq:bounded identity map}), and $W$ is stochastic.
Assume that $x^*$ is the unique equilibrium.
Suppose on the contrary that for all $j \in N$ we have $x^*_{j} = \sum_i x^*_{i} W_{ij} + \e_j < u_j$ and $x^*_{j} = \sum_i x^*_{i} W_{ij} + \e_j > \ell_j$, whence
$\ell \ll x^*W + \e \ll u$ and $x^* = f(x^*) = x^* W + \e$.
Since $W$ is stochastic and hence non-negative, Perron-Frobenius theorem implies that $W$ has a non-negative left eigenvector $v$ with respect to the eigenvalue $1$. That is, $v W = v$.
Now, since $\ell \ll x^* \ll u$, we can find $\l \neq 0$  such that $\ell \leq x^* + \l v \leq  u $.
Then, since $(x^* + \l v ) W + \e = x^* W + \l v + \e  =  x^* + \l v$, we have  $\ell \leq (x^* + \l v) W + \e \leq u$.
Therefore, we get $f((x^* + \l v )W + \e) = (x^* + \l v )$, so that $(x^* + \l v )$ is another equilibrium.
Since $x^*$ is the unique equilibrium, we have a contradiction.
\end{proof}

\begin{proof}[Proof of Lemma \ref{lemma:algorithm n2^n}]
Let $(f, W, \e)$ follows the conditions of lemma.
Let $u=(u_i)$ and $\ell=(\ell_i)$ be the upper bounds and lower bounds of $f$, respectively.
Given the states $x \in [\ell, u]$, let $A(x)$ and $B(x)$ be the sets of agents with the equilibrium at the bounds as (\ref{eq:sets of agents at boundedness}).
Denote $x^*$ as the unique equilibrium.
Let $A^* \coloneqq A(x^*) = \{j \in N \colon \sum_{i}x^*_i w_{ij} + \e_j \geq u_j \}$ be the sets of agents having the equilibrium at the upper bounds, and $B^* \coloneqq B(x^*) =  \{j \in N \colon \sum_{i}x^*_i w_{ij} + \e_j \leq \ell_j \}$ be the sets of agents with the equilibrium at the lower bounds.
Following the algorithm, if $x^* = u$ or $x^*=\ell$, then step $2$ or step $3$ of Algorithm \ref{algorithm} terminates and returns the equilibrium.
Without loss of generality, suppose that $x^* \neq u, \ell$.

Define $\mathcal{P} \coloneqq \mathcal{P}(B(\ell))$ as the power set of $B(\ell)$.
Then, since $B^* \subset B(\ell) $, there exists $P \in \mathcal{P}$ such that $P = B^*$. 
Define $\hat{u} \coloneqq \ell \Lambda^{B^*} + u (I - \Lambda^{B^*})$. 
Hence, economic states $\hat{u}$ indicate that an agent $i$ in $B^*$ has economic state $\ell_i$, and an agent $i$ in $N\setminus B^*$ has state $u_i$.

Since $x^* > \ell_j$ for $j \in N\setminus B^*$, we can be reduced the system to:
\begin{equation}\label{eq:algorithm:reduced system}
    \begin{split}
    \forall j \in B^*, \qquad  x_j & = \ell_j\\
    \forall j \in N\setminus B^*,  \quad x_j & =\min\left\{\max\left\{\sum_{i \in N\setminus B^*} x_i w_{ij}+ \sum_{i \in B^*} \ell_i w_{ij}+ \e_j, \ell_j \right\}, u_j \right\}\\
        & = \min\left\{\sum_{i \in N\setminus B^*} x_i w_{ij} + \sum_{i \in B^*} \ell_i w_{ij}+ \e_j, u_j \right\} \\
        &> \ell_j
    \end{split}
\end{equation}

We write (\ref{eq:algorithm:reduced system}) in vector form and define the operator $\hat{T}:[\ell, \hat{u}] \ra [\ell, \hat{u}]$ as 
\begin{equation*}
    x = \hat{T} x \coloneqq \{[x (I - \Lambda^{B^*}) W + \ell \Lambda^{B^*}W + \e ](I - \Lambda^{B^*})\}\land  [u (I - \Lambda^{B^*})] + \ell \Lambda^{B^*}.
\end{equation*}
Then, $x^*$ is the fixed point of $\hat{T}$.
Following the algorithm, we let $x^{(0)} = \hat{u}$ and $A_0 \coloneqq A(\hat{u}) \cap (N\setminus B^*) $, so that $A^* \subset A_0  \subset N\setminus B^*$.
Define $x^{(t)}$ for $t \in \N$ as the fixed point of $\Phi_t$:
\begin{multline}\label{eq:Phi_t}
   \Phi_t x \coloneqq u \Lambda^{A_{t-1}} + \ell \Lambda^{B^*} \\ + [u \Lambda^{A_{t-1}} W + \ell \Lambda^{B^*} W +  x (I -\Lambda^{A_{t-1}} -  \Lambda^{B^*}) W + \e](I -\Lambda^{A_{t-1}} -  \Lambda^{B^*}). 
\end{multline}
We have $x^{(t)} = \Phi_t x^{(t)}$ for all $t \in \N$. 
For $t \in \N$, define  $A_t = A(x^{(t)})$.
We first show that $\hat{T} x^{(t)} \leq x^{(t)}$ and $x^{(t+1)} \leq x^{(t)}$ for all $t=0, 1, 2, \dots$ by induction.

Consider $t=0$. Clearly, we have $\hat{T} x^{(0)} \leq x^{(0)}$.
We need to check that $\Phi_1$ has a unique fixed point.
Since $A_0 \supset A^*$, the matrix $I-(I - \Lambda^{A_0} - \Lambda^{B^*})W (I - \Lambda^{A_0} - \Lambda^{B^*})$ is non-singular.
Otherwise, there exists an irreducible submatrix $W_s$ of $(I - \Lambda^{A_0} - \Lambda^{B^*})W (I - \Lambda^{A_0} - \Lambda^{B^*})$ such that $r(W_s)=1$.
But, using Lemma \ref{lemma:irreducible, generically unique}, the proof of Proposition \ref{proposition:generic uniqueness and uniqueness}, and Lemma \ref{lemma:unique solution at boundedness}, since the equilibrium is unique, every strongly connected subgraph must contain a node $m$ such that its equilibrium is at the bound.
Since the operation $(I - \Lambda^{A_0} - \Lambda^{B^*})W (I - \Lambda^{A_0} - \Lambda^{B^*})$ removes such node $m$ from the graph $W$, Lemma \ref{lemma:submatrix r(W)<1} implies that $r(W_s) < 1$.
Therefore, such irreducible submatrix $W_s$ with $r(W_s)=1$ does not exist, so the spectral radius of $(I - \Lambda^{A_0} - \Lambda^{B^*})W (I - \Lambda^{A_0} - \Lambda^{B^*})$ is less than one.
Therefore, the fixed point of $\Phi_1$ is unique, and we can compute the fixed point $x^{(1)}$ by the iteration from any initial guess.
In particular, we can iterate from $x^{(0)}$: $x^{(1)} = \lim_{m \ra \infty} \Phi_1^m(x^{(0)})$.
Moreover, since $x^{(0)} =u (I - \Lambda^{B^*}) + \ell \Lambda^{B^*} $ and then $x^{(0)} (I -\Lambda^{A_{0}} -  \Lambda^{B^*}) = u (I -\Lambda^{A_{0}} -  \Lambda^{B^*})$, we have
\begin{equation*}
    \begin{split}
         \Phi_1 (x^{(0)})  & = u \Lambda^{A_{0}} + \ell \Lambda^{B^*}  \\
         & \quad + [u \Lambda^{A_{0}} W + \ell \Lambda^{B^*} W +  x^{(0)} (I -\Lambda^{A_{0}} -  \Lambda^{B^*}) W + \e](I -\Lambda^{A_{0}} -  \Lambda^{B^*}) \\
         & =  u \Lambda^{A_{0}} + \ell \Lambda^{B^*}
         + [u(I - \Lambda^{B^*})W   + \ell \Lambda^{B^*} W + \e ](I -\Lambda^{A_{0}} -  \Lambda^{B^*})\\
         & \leq u \Lambda^{A_{0}} + \ell \Lambda^{B^*} + u (I -\Lambda^{A_{0}} -  \Lambda^{B^*})\\
         &= u (I - \Lambda^{B^*}) + \ell \Lambda^{B^*} = x^{(0)}
    \end{split}
\end{equation*}
where the inequality holds because the definition of $A_0$ implies that if $A_0 \subsetneq N \setminus B^*$, then $ \sum_{i \in N \setminus B_{0}} u_i w_{ij} + \sum_{i \in B_{0}} \ell_i w_{ij} +  \e_j < u_j$ for $j \notin A_0$; otherwise, if $A_0 = N \setminus B^*$, then $I -\Lambda^{A_{0}} -  \Lambda^{B^*} = 0$ and $x^{(0)}=\hat{u}$ is the fixed point of $\hat{T}$.
Since  $\Phi_t$ is an increasing operator for all $t$, we get $\Phi_1^m(x^{(0)}) \leq \Phi_1^{m-1}(x^{(0)}) \leq \cdots \leq \Phi_1 (x^{(0)}) \leq x^{(0)}$ for all $m \in \N$, so that $x^{(1)} = \lim_{m \ra \infty} \Phi_0^m(x^{(0)}) \leq x^{(0)}$.

Suppose that $\hat{T} x^{(k)} \leq x^{(k)}$ holds for some $t = k \in \N$.
Since $A_{k} = A(x^{(k)}) = \{j \in N: \sum_{i} x^{(k)} w_{ij} + \e_j \geq u_j\}$, 
we have $x^{(k)} (I - \Lambda^{A_k} - \Lambda^{B^*}) + u  \Lambda^{A_k} + \ell\Lambda^{B^*} = x^{(k)}$.
Hence, we have 
\begin{equation}\label{eq:x^k remains constant}
    \begin{split}
        \Phi_{k+1} (x^{(k)}) & =  u \Lambda^{A_k} + \ell \Lambda^{B^*} \\
        & \ + [u \Lambda^{A_{k}} W + \ell \Lambda^{B^*} W +  x^{(k)} (I -\Lambda^{A_{k}} -  \Lambda^{B^*}) W + \e](I -\Lambda^{A_{k}} -  \Lambda^{B^*}) \\
        &= u \Lambda^{A_k} + \ell \Lambda^{B^*} + (x^{(k)} W + \e ) (I -\Lambda^{A_{k}} -  \Lambda^{B^*}) \\
        & = \hat{T} x^{(k)} \leq x^{(k)}
    \end{split}
\end{equation}
Similar to the above argument, $I - (I -\Lambda^{A_{k}} -  \Lambda^{B^*}) W (I -\Lambda^{A_{k}} -  \Lambda^{B^*})$ is non-singular, so $x^{(k+1)}$ is unique.
Then, since $\Phi_{k+1}$ is an increasing operator, and its iteration converges from any initial point, we have $x^{(k+1)} = \lim_{m \ra \infty }\Phi_{k+1}^m (x^{(k)}) \leq \cdots \leq \Phi_{k+1}(x^{(k)}) \leq x^{(k)}$.
Next, since $x^{(k+1)} \leq x^{(k)} $, we have $A_{k+1} = A(x^{(k+1)} ) \subset A(x^{(k)} ) = A_k$.
If $A_{k+1} = A_k$, then using the definition of $A_{k+1}$ we have 
\begin{equation*}
\begin{split}
\hat{T} x^{(k+1)} & = \{[x^{(k+1)} (I - \Lambda^{B^*}) W + \ell \Lambda^{B^*}W + \e ](I - \Lambda^{ B^*})\}\land  [ u (I - \Lambda^{B^*})] + \ell \Lambda^{B^*}  \\
 & = [x^{(k+1)} (I - \Lambda^{B^*} - \Lambda^{A_{k+1}}) W + u \Lambda^{A_{k+1}} W+  \ell \Lambda^{B^*}W + \e ](I - \Lambda^{ B^*})(I - \Lambda^{A_{k+1}}) \\
 & \quad + u \Lambda^{A_{k+1}} + \ell \Lambda^{B^*}\\
 & =  [x^{(k+1)} (I - \Lambda^{B^*} - \Lambda^{A_{k}}) W + u \Lambda^{A_{k}} W+  \ell \Lambda^{B^*}W + \e ](I - \Lambda^{ B^*} - \Lambda^{A_{k}}) \\
 & \quad + u \Lambda^{A_{k}} + \ell \Lambda^{B^*}\\
 & = \Phi_{k+1} x^{(k+1)} = x^{(k+1)} \\
\end{split}
\end{equation*}

Then, $x^{(k+1)}$ is the fixed point of $\hat{T}$ if $A_{k+1} = A_k$.
Moreover,  since $A_{k+1} = A_k$ implies $\Phi_{k+1} = \Phi_{k+2}$, we have $x^{(k+1)} = x^{(k+2)}$.

On the other hand, suppose that $A_{k+1} \subsetneq A_k$.
Denote $\hat{T} x^{(k+1)}_j$ and $\Phi_{k+1} x^{(k+1)}_j$ as the $j$-th entry of $\hat{T} x^{(k+1)}$ and $\Phi_{k+1} x^{(k+1)}$, respectively.
Then, for $j \in A_{k+1}$, we have $\hat{T} x^{(k+1)}_j = u_j = \Phi_{k+1} x^{(k+1)}_j = x^{(k+1)}_j$, where the second equality holds because $j \in A_k$.
For $j \in A_k \setminus A_{k+1}$, we have $\hat{T} x^{(k+1)}_j < u_j =  \Phi_{k+1} x^{(k+1)}_j = x^{(k+1)}_j$.
For $j \in B^*$, we have $\hat{T}x^{(k+1)}_j = \ell_j = \Phi_{k+1} x^{(k+1)}_j = x^{(k+1)}_j$.
For $j \in N \setminus (A_k \cup B^*)$, we have $j \notin (A_{k+1} \cup B^*)$ so that 
\begin{align*}
    \hat{T} x^{(k+1)}_j &= \sum_{i\in B^*}\ell_i w_{ij} +  \sum_{i \in A_{k+1}}  u_i w_{ij}  + \sum_{i \in N \setminus (A_{k+1} \cup B^*)} x^{(k+1)}_i w_{ij} + \e_j \\
    & = \sum_{i\in B^*}\ell_i w_{ij} +  \sum_{i \in A_{k+1}}  u_i w_{ij}  + \sum_{i \in N \setminus (A_{k} \cup B^*)} x^{(k+1)}_i w_{ij} + \sum_{i \in A_k \setminus A_{k+1}} x^{(k+1)}_i w_{ij} + \e_j \\
    & = \sum_{i\in B^*}\ell_i w_{ij} +  \sum_{i \in A_{k+1}}  u_i w_{ij}  + \sum_{i \in N \setminus (A_{k} \cup B^*)} x^{(k+1)}_i w_{ij} + \sum_{i \in A_k \setminus A_{k+1}} u_i w_{ij} + \e_j \\
    & = \sum_{i\in B^*}\ell_i w_{ij} +  \sum_{i \in A_{k}}  u_i w_{ij}  + \sum_{i \in N \setminus (A_{k} \cup B^*)} x^{(k+1)}_i w_{ij}  + \e_j \\
    & = \Phi_{k+1} x^{(k+1)}_j = x^{(k+1)}_j 
\end{align*}
where the inequality holds because $x_i^{(k+1)} = u_i$ for $i \in A_{k}$ by (\ref{eq:Phi_t}).
Therefore, we see that $\hat{T} x^{(k+1)} \leq x^{(k+1)}$ when $A_{k+1} \subsetneq A_k$.
Then, we get $\hat{T} x^{(k+1)} \leq x^{(k+1)}$.
The Mathematical Induction shows that $\hat{T} x^{(t)} \leq x^{(t)}$ and $x^{(t+1)} \leq x^{(t)}$ for all $t = 0, 1, 2, \dots$.

The above argument also shows that if $A_{k+1} = A_k$ for some $k >0$, then $x^{(k+1)}$ is the fixed point of $\hat{T}$.
Also, (\ref{eq:x^k remains constant}) implies that $x^{(k+2)}=\Phi_{k+2}x^{(k+1)} = \hat{T} x^{(k+1)} = x^{(k+1)}$, so that $x^{(t)}$ for $t \geq k+1$ remains constant.
On the other hand, if $x^{(k+1)}$ is not the fixed point of $\hat{T}$, since $x^{(k+1)}$ is the fixed point of $\Phi_{k+1}$, there exists $j \in A_{k} \setminus A_{k+1}$ such that $A_{k+1}$ decreases.
Since there are only $n$ nodes, and $A_0$ contains at most $n-1$ elements due to $x^* \neq u$, $A_t$ and $x^{(t)}$ stop to change after at most $n$ iterations.
Since the sequence $x^{(t)}$ is constant only at fixed point by (\ref{eq:x^k remains constant}), we obtain the equilibrium in at most $n$ iterations.
Now, in searching of $P=B^*$ from the power set of $B(\ell)$, since $x^* \neq \ell$ and then $B(\ell)$ contains at most $n-1$ agents, $\mathcal{P}$ has at most $2^{n-1}$ elements.
Then, we need at most $2^{n-1}$ searching time for $P=B^*$ and $n$ iterations for each possible $P \in \mathcal{P}$ to find the fixed point, so the overall iteration is at most $n 2^{n-1}$.
\end{proof}
\begin{proof}[Proof of Lemma \ref{lemma:linear programming}]
Let $f$ follow (\ref{eq:bounded identity map}), $W \geq 0$ be stochastic, and $\e \in \R^n$.
Let $g\colon \R^n \ra R$ be a strictly increasing function.
Suppose that $x^*$ is a solution to programming problem (\ref{linear programming}).
Then, $x^*$ satisfies $x^* \in [\ell, u]$ and $x^* \leq \max\{x^* W + \e , \ell \}$.
We want to show that $x^*_j$ must satisfies either $x^*_j = u_j$ or $x^*_j =(\max\{x^* W + \e , \ell  \})_j$ for all $j \in \N$, where $(\max\{x^* W + \e, \ell \})_j$ is the $j$-th entry of $\max\{x^* W + \e, \ell \}$.
Suppose not. 
Then, there is $i \in N$ such that $x^*_i < u_i$ and $x^*_i < (\max\{x^* W + \e, \ell \})_i$.
Then, we can find $\d > 0$ such that $x^*_i + \d \leq u_i$ and
$$x^*_i + \d \leq  (\max\{x^* W + \e, \ell \})_i \leq (\max\{(x^* + \d \mathbbm{1}^{i})W + \e, \ell \})_i,$$
where $\mathbbm{1}^{i}_j = 1$ if $j=i$ and $\mathbbm{1}^{i}_j = 0$ otherwise.
Therefore, $x^* + \d \mathbbm{1}^{i}$ satisfies the conditions of programming problem (\ref{linear programming}).
Since $g$ is strictly increasing, we have $g(x^*) < g(x^* + \d \mathbbm{1}^{i})$, contradicting that $x^*$ is the solution.
Hence, it must be that either $x^*_j = u_j$ or $x^*_j =(\max\{x^* W + \e , \ell  \})_j$ for all $j \in \N$, so that $x^* = \min\{\max\{x^* W + \e , \ell  \} , u\}$.
\end{proof}

\section{Proofs in Section \ref{sec:key player}}

\begin{proof}[Proof of Lemma \ref{lemma:continuous dynamics stable}]
Suppose that $f_i$ is increasing and continuously differentiable for all $i$, and $r(|W|\diag \b) < 1$.
Let $F(x) \coloneqq f(xW + \e) - x$. Then, the system (\ref{eq:continuous dynamics}) is $\de x / \de t = F(x)$.
We follow Lyapunov’s linearization method.
We have the partial derivatives
\begin{equation*}
    \frac{\p F_i(x)}{\p x_m} = 
    \begin{cases}
    f_i' (\sum_h x_h w_{hi} + \e_i) w_{ii} - 1 & \text{ if $m = i$,} \\
    f_i' (\sum_h x_h w_{hi} + \e_i) w_{mi} & \text{ if $m \neq i$,}
    \end{cases}
\end{equation*}

Denote $f' \coloneqq (f'_1(\sum_h x_h w_{h1} + \e_1), \dots,  f'_n(\sum_h x_h w_{hn} + \e_n))$.
The Jacobian matrix of $F$ is $ W \diag(f')- I$.
Let $\lambda$ be the eigenvalue of $W \diag(f')- I$.
The dynamics is asymptotically stable if the real part of $\lambda$, denoted as $\text{Re}(\lambda)$, is negative for all $\lambda$.
From the characteristic equation, eigenvalue $\lambda$ satisfies $0 = \det((W \diag(f')- I) - \lambda I) = \det ((W \diag(f') -(\lambda + 1) I ) $, so $\lambda + 1$ is the eigenvalue of $W \diag(f')$.
We then have
$\text{Re}(\lambda) + 1 \leq |\lambda + 1|  \leq r(W \diag(f'))$.

Since the interaction functions are Lipschitz continuous, the derivatives are bounded $|f'_i| \leq \b_i$ for all $i$, so that $\diag(f') \leq \diag(\b)$.
Then, we can show that $|(W \diag(f'))^k| \leq (|W| \diag(\b))^k$  for $k \geq 1$.
Since $r(A) \leq \|A^k\|^{1/k}$ for a matrix $A \in \R^{n \times n}$ and $k\geq 1$, we have $r(W \diag(f')) \leq \| (W \diag(f'))^k\|^{1/k} \leq  \| (|W| \diag(\b))^k\|^{1/k}$ for all $k \geq 1$.
Taking the limit of $k$, we have $\text{Re}(\lambda) \leq r(W \diag(f')) - 1 \leq r(|W| \diag(\b)) - 1 < 0$.
Since the real part of any eigenvalue of $W\diag(f')$ is negative, the system is asymptotically stable.
\end{proof}

\begin{proof}[Proof of Lemma \ref{lemma:key player measure}]
Let $f_i$ be differentiable for all $i$ and $r(|W|\diag \b) < 1$.
Define the system and function $F$ as (\ref{eq:continuous dynamics shocks}).
Define $C_{ij}$ as
$$C_{ij} \coloneqq \frac{\de x_j^*}{\de s_i}\frac{\p s_i}{\p x_i^*} x_i^*$$
for $i, j \in N$.
For the steady state, by (\ref{eq:continuous dynamics shocks}), we have $x_i^* = f(\sum_h x_h^* w_{hi} + \e_i)/s_i $, so that 
$$\p x_i^* / \p s_i = - f\lp\sum_h x_h^* w_{hi} + \e_i \rp / s_i^2 = - x^*_i / s_i.$$
For $s = \bm 1$, we have 
\begin{equation}\label{eq:C_ij}
  C_{ij} = \frac{\de x_j^*}{\de s_i}\frac{\p s_i}{\p x_i^*} x_i^* = \frac{\de x_j^*}{\de s_i} \lp\frac{- s_i}{x_i^*}\rp x_i^* = -\frac{\de x_j^*}{\de s_i}.  
\end{equation}
After the removal of $i$, the system goes to a new steady state near the original one.
Since $F=0$ at both new and original steady states, we have $\de F / \de s = \bm 0$, where $\bm 0$ is an $n$ by  $n$ zero matrix.
The total derivative $\de F = (\p F/ \p s) \de s + (\p F/ \p x) \de x $ gives
\begin{equation} \label{eq:total derivative}
  \frac{\de F}{\de s} = \frac{\p F}{\p s} + \frac{\p F}{\p x}\frac{\de x}{\de s} = \bm 0 .  
\end{equation}
By (\ref{eq:C_ij}) and (\ref{eq:total derivative}), since $\p f/\p x = W \diag({f'})$, we have 
\begin{equation*}
    \begin{split}
    C &= - \frac{\de x^*}{\de s} = - \lb \left. \frac{\p F}{\p s} \right\rvert_{(x, s) = (x^*, \bm 1)}\rb \lb \left. \frac{\p F}{\p x}\right\rvert_{(x, s) = (x^*, \bm 1)} \rb^{-1}  \\
    & = \diag{(x^*)} \lb I - W \diag{(f'(x^* W + \e))}\rb^{-1} 
    \end{split}
\end{equation*}
where $f'(x^* W + \e) =(f_1'(\sum_{h} x^*_h w_{h1} \e_1), \dots,  f_n'(\sum_{h} x^*_h w_{h1} \e_n))$.
The result follows from that $\sigma = (C \bm 1 ^\top)^\top $.
\end{proof}

\section{Other Lemmas}\label{sec:other lemmas}

\begin{lemma}\label{lemma:non-negative I-W inverse}
Let $A \in \R^{n \times n}$ be non-negative.
Then, $r(A) < 1$ if and only if $(I-A)^{-1}$ exists and $(I-A)^{-1} \geq 0$.
\end{lemma}

Let $A \in \R^{n \times n}$ be a matrix. Define $A_{-i}$ as the submatrix of $A$ deleting row $i$ and column $i$ of $A$.

\begin{lemma}
Let $A \in \R^{n \times n}$ be a non-negative matrix. Then, $r(A_{-i}) \leq r(A)$.
\end{lemma}

\begin{lemma}\label{lemma:submatrix r(W)<1}
Let $A \in \R^{n \times n}$ be an irreducible stochastic matrix. Then, $r(A_{-i}) < 1$.
\end{lemma}
\begin{proof}
Let $A=(a_{ij}) \in \R^{n \times n}$ be an irreducible stochastic matrix.
It is enough to consider the strongly connected graph with the least number of links, the ring.
That is, consider the graph such that for all $i \in N$ we have $a_{i, j}=1$ if $j = (i+1)\pmod{n}$, otherwise $a_{i, j}=0$.
Fix $i \in N$ and delete row $i$ and column $i$ to get $A_{-i}$.
Then, the row sum of row $i-1$ is less than one. 
Also, every other node has a path to node $i-1$.
Therefore, $A_{-i}$ is weakly chained substochastic, so it follows from Lemma \ref{condition:convergence} that $r(A_{-i}) < 1$.
We can extend the argument to an irreducible matrix, since if we remove a node $i$ from a strongly connected graph, then $i$ must be in a ring, and any other nodes have some path to that ring.
\end{proof}

%% file: main.bbl
\begin{thebibliography}{48}
\newcommand{\enquote}[1]{``#1''}
\expandafter\ifx\csname natexlab\endcsname\relax\def\natexlab#1{#1}\fi

\bibitem[\protect\citeauthoryear{Acemoglu, Akcigit, and Kerr}{Acemoglu
  et~al.}{2016{\natexlab{a}}}]{Production:acemoglu_Akcigit_Kerr_2016}
\textsc{Acemoglu, D., U.~Akcigit, and W.~Kerr} (2016{\natexlab{a}}):
  \enquote{Networks and the macroeconomy: An empirical exploration,} \emph{Nber
  macroeconomics annual}, 30, 273--335.

\bibitem[\protect\citeauthoryear{Acemoglu and Azar}{Acemoglu and
  Azar}{2020}]{production:acemoglu2020endogenous}
\textsc{Acemoglu, D. and P.~D. Azar} (2020): \enquote{Endogenous production
  networks,} \emph{Econometrica}, 88, 33--82.

\bibitem[\protect\citeauthoryear{Acemoglu, Carvalho, Ozdaglar, and
  Tahbaz-Salehi}{Acemoglu et~al.}{2012}]{Production:acemoglu2012network}
\textsc{Acemoglu, D., V.~M. Carvalho, A.~Ozdaglar, and A.~Tahbaz-Salehi}
  (2012): \enquote{The network origins of aggregate fluctuations,}
  \emph{Econometrica}, 80, 1977--2016.

\bibitem[\protect\citeauthoryear{Acemoglu, Ozdaglar, and
  Tahbaz-Salehi}{Acemoglu et~al.}{2015{\natexlab{a}}}]{acemoglu2015systemic}
\textsc{Acemoglu, D., A.~Ozdaglar, and A.~Tahbaz-Salehi} (2015{\natexlab{a}}):
  \enquote{Systemic risk and stability in financial networks,} \emph{American
  Economic Review}, 105, 564--608.

\bibitem[\protect\citeauthoryear{Acemoglu, Ozdaglar, and
  Tahbaz-Salehi}{Acemoglu et~al.}{2015{\natexlab{b}}}]{acemoglu2015Endogenous}
---\hspace{-.1pt}---\hspace{-.1pt}--- (2015{\natexlab{b}}): \enquote{Systemic
  risk in endogenous financial networks,} \emph{Columbia business school
  research paper}.

\bibitem[\protect\citeauthoryear{Acemoglu, Ozdaglar, and
  Tahbaz-Salehi}{Acemoglu et~al.}{2016{\natexlab{b}}}]{acemoglu2016network}
---\hspace{-.1pt}---\hspace{-.1pt}--- (2016{\natexlab{b}}): \enquote{Networks,
  Shocks, and Systemic Risk,} in \emph{The Oxford Handbook of the Economics of
  Networks}, ed. by Y.~Bramoullé, A.~Galeotti, and B.~W. Rogers, Oxford
  University Press.

\bibitem[\protect\citeauthoryear{Acemoglu, Ozdaglar, and
  Tahbaz-Salehi}{Acemoglu et~al.}{2017}]{production:acemoglu2017microeconomic}
---\hspace{-.1pt}---\hspace{-.1pt}--- (2017): \enquote{Microeconomic origins of
  macroeconomic tail risks,} \emph{American Economic Review}, 107, 54--108.

\bibitem[\protect\citeauthoryear{Amini, Filipovi{\'c}, and Minca}{Amini
  et~al.}{2016}]{financial:amini2016uniqueness}
\textsc{Amini, H., D.~Filipovi{\'c}, and A.~Minca} (2016): \enquote{Uniqueness
  of equilibrium in a payment system with liquidation costs,} \emph{Operations
  Research Letters}, 44, 1--5.

\bibitem[\protect\citeauthoryear{Antr{\`a}s, Chor, Fally, and
  Hillberry}{Antr{\`a}s et~al.}{2012}]{antras2012measuring}
\textsc{Antr{\`a}s, P., D.~Chor, T.~Fally, and R.~Hillberry} (2012):
  \enquote{Measuring the upstreamness of production and trade flows,}
  \emph{American Economic Review}, 102, 412--16.

\bibitem[\protect\citeauthoryear{Azimzadeh}{Azimzadeh}{2019}]{azimzadeh2019contraction}
\textsc{Azimzadeh, P.} (2019): \enquote{A fast and stable test to check if a
  weakly diagonally dominant matrix is a nonsingular M-matrix,}
  \emph{Mathematics of Computation}, 88, 783--800.

\bibitem[\protect\citeauthoryear{Ballester, Calv{\'o}-Armengol, and
  Zenou}{Ballester et~al.}{2004}]{NetworkGame:ballester2004s:crime}
\textsc{Ballester, C., A.~Calv{\'o}-Armengol, and Y.~Zenou} (2004):
  \enquote{Who's who in crime network. wanted the key player,} Tech. rep., IUI
  Working Paper.

\bibitem[\protect\citeauthoryear{Ballester, Calv{\'o}-Armengol, and
  Zenou}{Ballester et~al.}{2006}]{networkgame:ballester2006s}
---\hspace{-.1pt}---\hspace{-.1pt}--- (2006): \enquote{Who's who in networks.
  Wanted: The key player,} \emph{Econometrica}, 74, 1403--1417.

\bibitem[\protect\citeauthoryear{Bartelme and Gorodnichenko}{Bartelme and
  Gorodnichenko}{2015}]{production:Bartelme_2015}
\textsc{Bartelme, D. and Y.~Gorodnichenko} (2015): \enquote{Linkages and
  Economic Development,} Working Paper 21251, National Bureau of Economic
  Research.

\bibitem[\protect\citeauthoryear{Berman and Plemmons}{Berman and
  Plemmons}{1994}]{berman1994nonnegative}
\textsc{Berman, A. and R.~J. Plemmons} (1994): \emph{Nonnegative matrices in
  the mathematical sciences}, SIAM.

\bibitem[\protect\citeauthoryear{Blume, Brock, Durlauf, and Jayaraman}{Blume
  et~al.}{2015}]{networkgame:blume2015linear}
\textsc{Blume, L.~E., W.~A. Brock, S.~N. Durlauf, and R.~Jayaraman} (2015):
  \enquote{Linear social interactions models,} \emph{Journal of Political
  Economy}, 123, 444--496.

\bibitem[\protect\citeauthoryear{Calv{\'o}-Armengol, Patacchini, and
  Zenou}{Calv{\'o}-Armengol et~al.}{2009}]{networkgame:calvo2009peer}
\textsc{Calv{\'o}-Armengol, A., E.~Patacchini, and Y.~Zenou} (2009):
  \enquote{Peer effects and social networks in education,} \emph{The review of
  economic studies}, 76, 1239--1267.

\bibitem[\protect\citeauthoryear{Carvalho}{Carvalho}{2008}]{production:carvalho2008}
\textsc{Carvalho, V.~M.} (2008): \emph{Aggregate fluctuations and the network
  structure of intersectoral trade}, The University of Chicago.

\bibitem[\protect\citeauthoryear{Carvalho}{Carvalho}{2014}]{production:carvalho2014micro}
---\hspace{-.1pt}---\hspace{-.1pt}--- (2014): \enquote{From micro to macro via
  production networks,} \emph{Journal of Economic Perspectives}, 28, 23--48.

\bibitem[\protect\citeauthoryear{Carvalho and Tahbaz-Salehi}{Carvalho and
  Tahbaz-Salehi}{2019}]{production:carvalho2019production}
\textsc{Carvalho, V.~M. and A.~Tahbaz-Salehi} (2019): \enquote{Production
  networks: A primer,} \emph{Annual Review of Economics}, 11, 635--663.

\bibitem[\protect\citeauthoryear{Cifuentes, Ferrucci, and Shin}{Cifuentes
  et~al.}{2005}]{Financial:cifuentes2005liquidity}
\textsc{Cifuentes, R., G.~Ferrucci, and H.~S. Shin} (2005): \enquote{Liquidity
  risk and contagion,} \emph{Journal of the European Economic Association}, 3,
  556--566.

\bibitem[\protect\citeauthoryear{Cohen-Cole, Patacchini, and Zenou}{Cohen-Cole
  et~al.}{2015}]{networkgame:cohen2015static}
\textsc{Cohen-Cole, E., E.~Patacchini, and Y.~Zenou} (2015): \enquote{Static
  and dynamic networks in interbank markets,} \emph{Network Science}, 3,
  98--123.

\bibitem[\protect\citeauthoryear{Das, Samanta, and Pal}{Das
  et~al.}{2018}]{das2018study}
\textsc{Das, K., S.~Samanta, and M.~Pal} (2018): \enquote{Study on centrality
  measures in social networks: a survey,} \emph{Social network analysis and
  mining}, 8, 1--11.

\bibitem[\protect\citeauthoryear{Eisenberg and Noe}{Eisenberg and
  Noe}{2001}]{eisenberg2001systemic}
\textsc{Eisenberg, L. and T.~H. Noe} (2001): \enquote{Systemic risk in
  financial systems,} \emph{Management Science}, 47, 236--249.

\bibitem[\protect\citeauthoryear{Elliott, Golub, and Jackson}{Elliott
  et~al.}{2014}]{elliott2014financial}
\textsc{Elliott, M., B.~Golub, and M.~O. Jackson} (2014): \enquote{Financial
  networks and contagion,} \emph{American Economic Review}, 104, 3115--53.

\bibitem[\protect\citeauthoryear{Elsinger, Lehar, and Summer}{Elsinger
  et~al.}{2006}]{financial:elsinger2006risk}
\textsc{Elsinger, H., A.~Lehar, and M.~Summer} (2006): \enquote{Risk assessment
  for banking systems,} \emph{Management science}, 52, 1301--1314.

\bibitem[\protect\citeauthoryear{Fletcher}{Fletcher}{1989}]{InputOutput:fletcher1989tourism}
\textsc{Fletcher, J.~E.} (1989): \enquote{Input-output analysis and tourism
  impact studies,} \emph{Annals of tourism research}, 16, 514--529.

\bibitem[\protect\citeauthoryear{Gai and Kapadia}{Gai and
  Kapadia}{2019}]{Financial:gai2019networks}
\textsc{Gai, P. and S.~Kapadia} (2019): \enquote{Networks and systemic risk in
  the financial system,} \emph{Oxford Review of Economic Policy}, 35, 586--613.

\bibitem[\protect\citeauthoryear{Galeotti, Golub, and Goyal}{Galeotti
  et~al.}{2020}]{networkgame:galeotti2020targeting}
\textsc{Galeotti, A., B.~Golub, and S.~Goyal} (2020): \enquote{Targeting
  interventions in networks,} \emph{Econometrica}, 88, 2445--2471.

\bibitem[\protect\citeauthoryear{Glasserman and Young}{Glasserman and
  Young}{2015}]{Financial:glasserman2015likely}
\textsc{Glasserman, P. and H.~P. Young} (2015): \enquote{How likely is
  contagion in financial networks?} \emph{Journal of Banking \& Finance}, 50,
  383--399.

\bibitem[\protect\citeauthoryear{Glasserman and Young}{Glasserman and
  Young}{2016}]{financial:glasserman2016contagion}
---\hspace{-.1pt}---\hspace{-.1pt}--- (2016): \enquote{Contagion in financial
  networks,} \emph{Journal of Economic Literature}, 54, 779--831.

\bibitem[\protect\citeauthoryear{Herskovic}{Herskovic}{2018}]{production:herskovic2018networks}
\textsc{Herskovic, B.} (2018): \enquote{Networks in production: Asset pricing
  implications,} \emph{The Journal of Finance}, 73, 1785--1818.

\bibitem[\protect\citeauthoryear{Herskovic, Kelly, Lustig, and
  Van~Nieuwerburgh}{Herskovic et~al.}{2020}]{production:herskovic2020firm}
\textsc{Herskovic, B., B.~Kelly, H.~Lustig, and S.~Van~Nieuwerburgh} (2020):
  \enquote{Firm volatility in granular networks,} \emph{Journal of Political
  Economy}, 128, 4097--4162.

\bibitem[\protect\citeauthoryear{Hurd}{Hurd}{2016}]{hurd2016contagion}
\textsc{Hurd, T.~R.} (2016): \emph{Contagion!: Systemic Risk in Financial
  Networks}, Springer.

\bibitem[\protect\citeauthoryear{Jackson and Pernoud}{Jackson and
  Pernoud}{2020}]{multiplicity:jackson2020credit}
\textsc{Jackson, M.~O. and A.~Pernoud} (2020): \enquote{Credit freezes,
  equilibrium multiplicity, and optimal bailouts in financial networks,}
  \emph{arXiv preprint arXiv:2012.12861}.

\bibitem[\protect\citeauthoryear{Liu, Paddrik, Yang, and Zhang}{Liu
  et~al.}{2020}]{Financial:liu2020interbank}
\textsc{Liu, A., M.~Paddrik, S.~Y. Yang, and X.~Zhang} (2020):
  \enquote{Interbank contagion: An agent-based model approach to endogenously
  formed networks,} \emph{Journal of Banking \& Finance}, 112, 105191.

\bibitem[\protect\citeauthoryear{Long and Plosser}{Long and
  Plosser}{1983}]{production:long1983real}
\textsc{Long, J.~B. and C.~I. Plosser} (1983): \enquote{Real business cycles,}
  \emph{Journal of political Economy}, 91, 39--69.

\bibitem[\protect\citeauthoryear{Miller and Blair}{Miller and
  Blair}{2009}]{miller2009input}
\textsc{Miller, R.~E. and P.~D. Blair} (2009): \emph{Input-output analysis:
  foundations and extensions}, Cambridge university press.

\bibitem[\protect\citeauthoryear{Miller and Temurshoev}{Miller and
  Temurshoev}{2017}]{miller2017Downstreamness_Upstreamness}
\textsc{Miller, R.~E. and U.~Temurshoev} (2017): \enquote{Output upstreamness
  and input downstreamness of industries/countries in world production,}
  \emph{International Regional Science Review}, 40, 443--475.

\bibitem[\protect\citeauthoryear{Pesaran and Yang}{Pesaran and
  Yang}{2020}]{production:pesaran2020econometric}
\textsc{Pesaran, M.~H. and C.~F. Yang} (2020): \enquote{Econometric analysis of
  production networks with dominant units,} \emph{Journal of Econometrics},
  219, 507--541.

\bibitem[\protect\citeauthoryear{Rogers and Veraart}{Rogers and
  Veraart}{2013}]{financial:rogers2013failure}
\textsc{Rogers, L.~C. and L.~A. Veraart} (2013): \enquote{Failure and rescue in
  an interbank network,} \emph{Management Science}, 59, 882--898.

\bibitem[\protect\citeauthoryear{Roukny, Battiston, and Stiglitz}{Roukny
  et~al.}{2018}]{multiplicity:roukny2018interconnectedness}
\textsc{Roukny, T., S.~Battiston, and J.~E. Stiglitz} (2018):
  \enquote{Interconnectedness as a source of uncertainty in systemic risk,}
  \emph{Journal of Financial Stability}, 35, 93--106.

\bibitem[\protect\citeauthoryear{Sharkey}{Sharkey}{2017}]{sharkey2017control}
\textsc{Sharkey, K.~J.} (2017): \enquote{A control analysis perspective on Katz
  centrality,} \emph{Scientific reports}, 7, 1--8.

\bibitem[\protect\citeauthoryear{Stachurski}{Stachurski}{2022}]{stachurski2022systemic}
\textsc{Stachurski, J.} (2022): \enquote{Systemic Risk in Financial Systems:
  Properties of Equilibria,} \emph{arXiv preprint arXiv:2202.11183}.

\bibitem[\protect\citeauthoryear{Staum, Feng, and Liu}{Staum
  et~al.}{2016}]{financial:staum2016systemic}
\textsc{Staum, J., M.~Feng, and M.~Liu} (2016): \enquote{Systemic risk
  components in a network model of contagion,} \emph{IIE Transactions}, 48,
  501--510.

\bibitem[\protect\citeauthoryear{Timmer, Dietzenbacher, Los, Stehrer, and
  De~Vries}{Timmer et~al.}{2015}]{InputOutputTable:timmer2015}
\textsc{Timmer, M.~P., E.~Dietzenbacher, B.~Los, R.~Stehrer, and G.~J.
  De~Vries} (2015): \enquote{An illustrated user guide to the world
  input--output database: the case of global automotive production,}
  \emph{Review of International Economics}, 23, 575--605.

\bibitem[\protect\citeauthoryear{Veraart}{Veraart}{2020}]{Financial:veraart2020distress}
\textsc{Veraart, L. A.~M.} (2020): \enquote{Distress and default contagion in
  financial networks,} \emph{Mathematical Finance}, 30, 705--737.

\bibitem[\protect\citeauthoryear{Zenou}{Zenou}{2012}]{networkgame:zenou2012networks}
\textsc{Zenou, Y.} (2012): \enquote{Networks in economics,} .

\bibitem[\protect\citeauthoryear{Zenou}{Zenou}{2016}]{NetworkGame:zenou2016key}
---\hspace{-.1pt}---\hspace{-.1pt}--- (2016): \enquote{Key players,}
  \emph{Oxford Handbook on the Economics of Networks}, 244--274.

\end{thebibliography}
